# Three-Dimensional Isotropic STED Nanoscopy using a Single Objective

Renlong Zhang[1, †], Xiaoyu Weng[1, †, *], Haoxian Zhou[1], Luwei Wang[1], Fangrui Lin[1], Wei Yan[1], Xiumin Gao[2], Bin Yu[1], Danying Lin[1], Liwei Liu[1], Chenshuang Zhang[1], Kayla K. Green[3], Ewoud R. E. Schmidt[3], Songlin Zhuang[2], Junle Qu[1, *]

*1 State Key Laboratory of Radio Frequency Heterogeneous Integration & Key Laboratory of Optoelectronic Devices and Systems, College of Physics and Optoelectronic Engineering, Shenzhen University, Shenzhen 518060, China*

*2 Engineering Research Center of Optical Instrument and System, Ministry of Education, Shanghai Key Lab of Modern Optical System, School of Optical-Electrical and Computer Engineering, University of Shanghai for Science and Technology, 516 Jungong Road, Shanghai 200093, China.*

*3 Department of Bioengineering, CU-MUSC Bioengineering Program, Clemson University, Charleston, South Carolina, 29634, USA*

†These authors contributed equally to this work.

* Correspondence and requests for materials should be addressed to X.W. (email: xiaoyu@szu.edu.cn ) or to J.Q. (email: jlqu@szu.edu.cn ).

# Abstract:

Accurate three-dimensional (3D) imaging requires an isotropic point spread function (PSF). However, the inherent missing aperture of a single objective lens results in an elongated, cigar-like PSF, which has rendered isotropic resolution in fluorescence microscopy seemingly insurmountable without a $4\pi$ configuration for decades. To address this long-standing challenge, we introduce ISO-STED (Isotropic Single-Objective STED) Nanoscopy, a novel approach that employs a single objective lens and a single depletion beam. By utilizing a hollow depletion focus, ISO-STED achieves an isotropic PSF without relying on a $4\pi$ configuration. This innovative design enables uniform fluorescence suppression in all directions, thereby yielding an isotropic 3D resolution of approximately 70 nm. Our work not only demonstrates the potential of ISO-STED Nanoscopy to provide a compact and versatile solution for isotropic 3D imaging in complex specimens but also paves the way for more accessible and practical applications in various research fields, including biomedical research and neuroscience.

# Introduction

Far-field optical microscopy has long served as a cornerstone technology across biology, medicine, and nanoscience, enabling direct observation of structures spanning from cellular to molecular scales [1-3]. However, its capabilities are fundamentally limited by two inherent constraints: the diffraction limit, which restricts spatial resolution, and the pronounced anisotropy between lateral and axial resolution inherent to single-objective architectures [4, 5]. Over past couple decades, super-resolution fluorescence microscopy such as stimulated emission depletion (STED), structured illumination microscopy (SIM), and single-molecule localization methods (PALM/STORM) have successfully surpassed the diffraction barrier in three dimensions [6-13]. Yet, these advances remain limited by the axially elongated point spread function (PSF) characteristic of single-objective systems, which arises from their inability to collect the complete spherical wavefront [4]. This intrinsic PSF asymmetry introduces geometric distortions in volumetric reconstructions, ultimately hindering accurate structural interpretation in three dimensions.

This challenge raises a fundamental question: can isotropic three dimensional (3D) super-resolution be achieved within the geometric constraints of a single-objective configuration? While long regarded as a physical limitation, most effective strategies for isotropic resolution to date have relied on dual-objective architectures, namely 4Pi microscopy [14]. By interfering counter-propagating wavefronts, these systems recover the missing half of the aperture and have been successfully integrated with super-resolution modalities, including 4pi-STED, 4pi-STORM, and 4pi-SIM [15-19]. Most demonstrate impressive performance in thin, transparent samples such as cultured cells. Among them, only 4pi-STED can achieve isotropic resolution down to ~50 nm in tissue [17]. However, 4Pi-based systems suffer from intrinsic drawbacks: their reliance on coherent interference makes them highly susceptible to refractive index mismatches and sample-induced aberrations, severely restricting imaging depth even with adaptive optics or tissue clearing [17, 18]. Furthermore, their dual-objective geometry imposes strict constraints on sample preparation, rendering them incompatible with thick, uneven, or in-vivo specimens [20, 21]. Undoubtedly, achieving isotropic super-resolution with a single objective is of great significance for nanoscale imaging in complex biological systems, but remains technically elusive.

Here, we introduce ISO-STED (Isotropic Single-Objective STED) Nanoscopy, a novel approach that employs a single objective lens and a single depletion beam. By integrating polarization mode extraction with precise engineering of focal energy distribution and axial positioning, ISO-STED generates a hollow depletion focus. This hollow depletion focus enables the achievement of an isotropic

PSF without relying on a 4π configuration. This innovative design facilitates uniform fluorescence suppression in all directions, thereby yielding an isotropic 3D resolution of approximately 70 nm. We demonstrate the versatility of ISO-STED across diverse biological systems, including the dynamics of mitochondrial cristae in live cells, layer-specific quantification of mitochondrial volumes in articular cartilage, and synaptic imaging at depths up to 30 μm in brain tissue. This approach allows for the direct transformation of a conventional 2D-STED system into an isotropic 3D-STED modality. By doing so, it paves the way for more accessible and practical applications in various research fields, including biomedical research and neuroscience.

## Results

### Theoretical Principle of ISO-STED

The realization of isotropic 3D super-resolution in a single-objective, single-depletion beam system hinges on fulfilling two essential criteria. First, it is imperative to create three focal points along the axial dimension, each characterized by precisely defined orbital angular momentum (OAM), intensity distribution, and spatial location. Second, to maintain the integrity of the 3D PSF, it is crucial to prevent crosstalk or interference among these foci by assigning them orthogonal polarization states. To meet these requirements, we have developed a combined strategy that integrates the optical pen technique and polarization mode extraction [22, 23]. The detailed implementation is provided in Supplementary Figure 1, as well as Supplementary Notes 1 and 2.

For the first condition, a 1-ns pulsed 775-nm laser beam is modulated by a phase-only SLM (spatial light modulator) in Fig. 1, producing three foci with topological charges of +m, –(m+1), and +m. The phase introduced by the optical pen can be written as [22]

$$\phi_{op} = Phase[\sum_{j=1}^{N} PF(s_j, x_j, y_j, z_j, \sigma_j, m_j)] \tag{1}$$

where $N$ is the number of foci defines, $x_j, y_j, z_j$ are the spatial position of the *j-th* focus, respectively. and $s_j$ and $\delta_j$ weight factors that can be used to adjust the amplitude and phase of the *j-th* focus, respectively. $m_j$ denotes the topological charge of the *j-th* focus. Therefore, optical pen indicated by Eq. (1) allow dynamic modulation of the number, position, and topological charges of the foci, see Supplementary Fig. 4.

For the second condition, while the aforementioned phase function of the optical pen enables controlled focal positioning, decoupling the polarization states of the beams is essential to prevent interference between adjacent foci. Specifically, a generalized linearly polarized light can be considered as the superposition of right- and left-circularly polarized modes. This relationship can be mathematically expressed as

$$E_m = \exp\left[imf\left(\varphi, \theta\right)\right]\left|\mathbf{R}\right\rangle + \exp\left[-imf\left(\varphi, \theta\right)\right]\left|\mathbf{L}\right\rangle \tag{2}$$

Here, $\left|\mathbf{R}\right\rangle = \begin{bmatrix} 1 & i \end{bmatrix}$ and $\left|\mathbf{L}\right\rangle = \begin{bmatrix} 1 & -i \end{bmatrix}$ denote the left right circularly polarized mode (RCP) and right circularly polarized mode (RCP), respectively. $\theta$ and $\varphi$ are the convergent and azimuthal angle,

respectively. In this paper, $f(\varphi,\theta)=\varphi$, and Equation (2) is simplified into a $m$-order vector vortex beam (VVB), which is created by the m-order vortex retarder [24, 25].

After modulation by $T=\exp in\varphi$, the $m$-order VVB in Equation (2) turn into

$$E_{cp}=\exp\left(i\left(m+n\right)\varphi\right)|\mathbf{R}\rangle+\exp\left(-i\left(m-n\right)\varphi\right)|\mathbf{L}\rangle \tag{3}$$

When $n=\pm m$, one can obtain

$$E_{mcL}=\exp\left(i2m\varphi\right)|\mathbf{R}\rangle+|\mathbf{L}\rangle \tag{4}$$

$$E_{mcR}=\exp\left(-i2m\varphi\right)|\mathbf{L}\rangle+|\mathbf{R}\rangle \tag{5}$$

From Eqs. (4, 5), the RCP and LCP with topological charge of $\pm 2m$ are always located at the outer ring, while the LCP and RCP modes with zero topological charge are positioned at the inner ring, as shown in Supplementary Figure 1(b). Generally, a larger topological charge results in a larger spatial separation and a lower energy density of the circularly polarized modes. Consequently, in the focal region of the objective, the lower circularly polarized modes with smaller topological charges can be effectively extracted from the $m$-order VVB.

As shown in Supplementary Figure 6, the order of VVB plays a critical role in ensuring adequate separation between these orthogonal components. When the topological charge $m \geq 3$, the main lobe (e.g., LCP) and the side ring lobe (RCP) are sufficiently separated spatially in the focal region of the objective. Moreover, when $m = 3$, the energy density of the outer polarized modes accounts for only 4.586% of the inner polarized modes. Because the depletion beam does not contribute to fluorescence excitation, and the outer polarized modes is both spatially displaced and energetically weak, its impact on imaging is negligible. For this reason, $m = 30$ in this paper, ensuring that the outer polarized modes have no measurable influence—either spatially or energetically on the inner desired polarization modes.

Based on the aforementioned polarization extraction technique and the use of an optical pen, two axially displaced foci (STEDz) with right-circular polarization state and zero topological charge, as well as a central focus (STEDxy) with left-circular polarization state and a topological charge of +1, are created in the focal region of the objective lens, as shown in Supplementary Figure 5. Notably, if the polarization states of STEDxy and STEDz are not rendered orthogonal, interference among the three foci leads to pronounced PSF distortion, as illustrated in Supplementary Figure 7. Therefore, ensuring polarization orthogonality is crucial for preventing interference between adjacent foci. By precisely tuning their axial positions and light intensity, a hollow, sphere-shaped depletion focus is ultimately formed using the objective lens, as shown in Supplementary Figure 4.

The inherent flexibility of ISO-STED in modulating focal intensity, spatial positioning, and polarization enables precise control over lateral and axial resolution. As demonstrated in Supplementary Figure 8, variations in focal position and relative intensity significantly impact the symmetry of the final PSF. In particular, improper polarization assignment leads to notable PSF distortions, compromising imaging resolution and structural fidelity. Furthermore, the relative energy distribution between $STED_{xy}$ and $STED_z$ is critical for achieving isotropic resolution. As shown in Supplementary Figure 9, gradually increasing the proportion of $STED_z$ power improves axial resolution, eventually matching the lateral resolution to yield a highly symmetric and isotropic PSF.

**ISO-STED optical system**

The phase pattern necessary for isotropic 3D depletion is composed of three main components (Fig. 1b): (1) two foci with *m*-order OAM are defocused at $z = \pm\Delta z$; (2) one central focus at the focal plane with OAM of –(m+1), the other two are *m*-order; and (3) a chromatic correction component. Due to the >100 nm wavelength difference between excitation (635 nm) and depletion (775 nm), axial chromatic aberration must be compensated. Accordingly, defocus values are adjusted across all three foci to ensure proper axial alignment.

To minimize power requirements while maximizing resolution, ISO-STED employs a 63× oil-immersion objective (NA = 1.4). The system utilizes 635 nm and 775 nm lasers (80 MHz repetition rate) compatible with standard STED dyes (the detailed information illustrated in Supplementary Figure 2). Simulations of the excitation beam, depletion beam, their spatial overlap, and the resulting effective PSF are presented in Fig. 1(c). Owing to its single-objective configuration, ISO-STED imposes no special constraints on sample geometry or preparation—enabling isotropic 3D super-resolution imaging of fixed cells, live cells, and thick tissue specimens.

To provide a clearer assessment, we quantified the system performance by imaging 149 crimson fluorescent beads with a diameter of 20 nm (Fig. 1(d) and Supplementary Figure 10). In confocal mode, the full-width at half-maximum (FWHM) of the PSF is measured to be 318 nm laterally and 760 nm axially. Using conventional 2D-STED (C-STED), the FWHM improves to 78 nm laterally and 501 nm axially. In contrast, ISO-STED achieves an isotropic resolution of 70 nm laterally and 72 nm axially. Statistical analysis of all 149 beads yields mean FWHMs of 73.4 nm laterally and 77.6 nm axially, confirming the system's capability for high-resolution, isotropic 3D imaging.

**ISO-STED enabled isotropic 3D super-resolution imaging of fixative cells**

To evaluate the imaging performance of ISO-STED, we image immunolabeled tubulin filaments in fixed HeLa cells (Fig. 2a, e, Supplementary Video 1). Compared to confocal microscopy, ISO-STED resolves fine filamentous structures that remain indistinct under diffraction-limited imaging, with both lateral and axial dimensions appearing equally well-defined. In contrast, C-STED enhances lateral resolution but fails to resolve closely spaced structures in the axial direction, exhibiting characteristic PSF elongation due to limited axial resolution (Fig. 2b, c). These results confirm the isotropic super-resolution capability of ISO-STED. We then image the nuclear pore complex protein Nup153 in fixed HeLa cells (Fig. 2g–i). ISO-STED not only resolves individual Nup153 puncta in the lateral dimension (Fig. 2k) but also distinctly reveals their double-layered axial distribution (Fig. 2h, i, Supplementary Video 2), which is poorly resolved using confocal and C-STED modalities.

To demonstrate the broader applicability of ISO-STED to other organelles, we image the outer mitochondrial membrane (OMM) in fixed HeLa cells (Supplementary Fig. 11). ISO-STED clearly resolves the hollow, reticulated structure of the mitochondrial network, while C-STED exhibits pronounced axial distortion and confocal microscopy fails to reveal the internal mitochondrial architecture. Finally, we leverage the 3D isotropic resolution of ISO-STED to investigate mitochondrial structural responses to oxidative stress. Cells treated with rotenone, a mitochondrial complex I inhibitor [26, 27], exhibit fragmented and disorganized mitochondrial networks with disrupted construction continuity, in contrast to the highly interconnected morphology observed in vehicle-treated controls.

**ISO-STED enabled dynamic super-resolution imaging in live cells**

To evaluate the potential of ISO-STED for live-cell imaging, we conduct isotropic 3D super-resolution imaging of the inner mitochondrial membrane (IMM) in live HeLa cells. Owing to its superior axial resolution and optical sectioning capabilities, ISO-STED effectively eliminates out-of-focus background fluorescence, thereby significantly enhancing image contrast compared with C-STED and confocal microscopy. As illustrated in Figure 3a–c, ISO-STED markedly outperforms both confocal and C-STED in removing defocused signals, providing a practical approach to improve image quality in two-dimensional super-resolution imaging. We further employ ISO-STED to visualize the 3D architecture of the inner mitochondrial membrane (IMM) (Fig. 3d (1–6)). While confocal microscopy fails to resolve any discernible substructure, ISO-STED unambiguously reveals the mitochondrial cristae. The limited axial resolution of confocal imaging results in pronounced z-axis elongation and morphological distortion, whereas ISO-STED accurately preserves the true mitochondrial geometry, including circular and ring-like cristae arrangements (Fig. 3e).

Leveraging the optical sectioning capabilities of ISO-STED, we monitor mitochondrial dynamics in live cells over a 63.15 s interval (Fig. 3f). Notably, we observe a transient transformation where the tip of a mitochondrion sharpens directionally and then retracts within approximately one second. Concurrently, the cristae within the same region undergo fragmentation and reorganization. These observations, documented in Supplementary Fig. 12 and Supplementary Video 3, provide representative examples of abrupt mitochondrial remodeling events.

**ISO-STED enabled imaging the thick tissue**

The single-objective design of ISO-STED imposes no special constraints on sample thickness, making it well-suited for volumetric imaging of thick biological tissues. To demonstrate this capability, we apply ISO-STED to image joint tissue sections from mice immunolabelled for the outer mitochondrial membrane (OMM). We acquire a large-volume dataset in confocal mode, spanning a depth of $30\,\mu m$ from the tissue surface ($z = 0$) to the interior ($z = 30\,\mu m$), with depth encoded by color (Fig. 4a, Supplementary Video 4). Within this $42.5 \times 42.5 \times 30\,\mu m^3$ imaging volume, we perform high-resolution ISO-STED imaging at two representative depths: $z = 4.8$–$7.2\,\mu m$ (Fig. 4b) and $z = 23.8$–$26.2\,\mu m$ (Fig. 4c), acquiring sub-volumes of $10.11 \times 10.11 \times 2.4\,\mu m^3$.

At both imaging depths, ISO-STED clearly resolves the hollow morphology of the mitochondrial outer membrane with high fidelity. In contrast, C-STED exhibits notable axial elongation and structural distortion, while confocal microscopy lacks the resolution to resolve any membrane features. As expected, increasing imaging depth introduces moderate degradation in resolution due to light scattering and absorption (Fig. 4d–g), with lateral resolution decreasing from 96 nm to 132 nm and axial resolution from 98 nm to 133 nm. Despite this degradation, the characteristic hollow structure of mitochondria remains clearly discernible, confirming that ISO-STED preserves isotropic resolution performance even in optically challenging environments with absorption and aberration.

To evaluate the generalizability of ISO-STED across tissue types, we image fixed mouse brain slices with fluorescently labelled neurons (Fig. 4h, Supplementary Video 5). We acquire a confocal scan of the entire $30\,\mu m$ depth range and then use ISO-STED to image two sub-regions at different depths (Fig. 4i–l). The isotropic resolution of ISO-STED enables accurate visualization of synaptic spines on dendritic processes, revealing detailed 3D morphology that would otherwise be inaccessible. While some resolution loss is observed with increasing depth, fine structural features remain clearly visible. Together, these results establish ISO-STED as a powerful and versatile platform for thick-tissue imaging.

Its compatibility with conventional STED sample preparation protocols and robust performance in diverse tissue environments make it well-suited for broad applications in complex biological specimens.

**ISO-STED reveals layer-specific mitochondrial differences in articular cartilage**

Articular cartilage is a structurally anisotropic tissue characterized by distinct zonal stratification along the depth axis, which reflects its specialized mechanical and biological functions. Based on collagen fiber orientation and chondrocyte organization, mouse cartilage can be subdivided into three major zones: the superficial zone, the middle zone, and the subchondral bone zone [28] (Fig. 5a). The superficial zone is enriched in horizontally flattened chondroprogenitor cells, which contribute to shear resistance and tissue regeneration [29, 30]. The middle zone comprises rounded chondrocytes typically arranged in pairs within a disorganized collagen network, forming a transition between the superficial and deep layers. In the subchondral bone zone, collagen fibers are vertically oriented to anchor the cartilage to the underlying bone, and cells are spherical, densely packed, and arranged randomly to resist compressive loads [31]. These structural characteristics are corroborated by hematoxylin and eosin (H&E) staining and confocal imaging (Fig. 5f–h).

While confocal imaging confirms the gross morphological differences among cartilage zones, we leverage the sub-100-nm isotropic resolution of ISO-STED to probe mitochondrial architecture at the subcellular level. Using FIJI's 3D Objects Counter plugin, we quantify individual mitochondrial volumes within each cartilage layer (Fig. 5b–c(3), Supplementary Fig. 13). Notably, cells in the subchondral bone zone harbour significantly larger mitochondria than those in the superficial and middle zones. The average mitochondrial volumes are $0.5142\ \mu m^3$ in the subchondral bone zone, $0.129\ \mu m^3$ in the middle zone, and $0.084\ \mu m^3$ in the superficial zone (Fig. 5e–h).

These results reveal that zonal differentiation in cartilage extends beyond tissue and cellular architecture to subcellular organelles. The volumetric differences in mitochondria may reflect distinct metabolic demands or oxidative stress responses across layers, underscoring the importance of mitochondrial structure as a functional biomarker in cartilage biology [28, 32]. By enabling nanoscale imaging in intact tissue, ISO-STED provides a powerful tool to uncover subtle organelle-level differences that may underpin regional vulnerability in cartilage-related diseases such as osteoarthritis.

## Discussion

A fundamental limitation in 3D optical nanoscopy is the anisotropic nature of the point spread function (PSF) in single-objective systems, which has long been considered a physical constraint imposed by incomplete wavefront collection. In this study, we reframe this physical limitation as an engineering challenge and introduce ISO-STED, a nanoscopy strategy that bridges the axial–lateral resolution gap using only a single objective and a single depletion beam. By employing coordinated spatial and polarization control, ISO-STED generates a hollow, spherically symmetric depletion focus, enabling 3D super-resolution imaging with approximately 70 nm isotropic resolution. This approach redefines the PSF design paradigm for STED nanoscopy and provides a broadly compatible platform for biological imaging in complex specimens.

A key innovation of ISO-STED is its ability to transform a conventional 2D-STED setup into a 3D isotropic system through single-pass phase modulation of a single depletion beam. Traditional 3D-STED approaches typically employ independent STEDxy and STEDz beams, requiring dual polarization control, multiple beam paths, or sequential modulation [33, 34]. ISO-STED overcomes these limitations by encoding multiple depletion foci with distinct axial positions, OAM states, energy distribution, and orthogonal polarization modes into a single hologram using a polarization mode extraction strategy. This design circumvents the symmetry constraints of $0$–$\pi$ phase masks, which offer limited control over foci positions and energy balance [12]. Instead, ISO-STED enables nanometer-scale, independent tuning of focal spacing and intensity through a single static phase profile, eliminating the need for power-based compensation (Supplementary Figs. 8–9). This flexibility not only optimizes PSF isotropy but also greatly simplifies system design and implementation.

Although raster-based volumetric scanning remains time-consuming and restricts temporal resolution, the strong axial confinement achieved in ISO-STED offers significant practical advantages. By efficiently suppressing out-of-focus fluorescence, ISO-STED enhances contrast and signal-to-noise ratio even in two-dimensional acquisitions (Fig. 3a). This property is critical for time-lapse imaging of mitochondrial cristae dynamics in live cells, enabling high-fidelity capture of sub-organelle structural remodeling. These results underscore ISO-STED's potential for dynamic live-cell imaging. Despite ongoing challenges such as photobleaching and phototoxicity, recent advances may further extend ISO-STED's utility for tracking fast physiological processes. These include the development of photostable STED-compatible fluorophores [35], bleach correction algorithms, high-speed scanning technologies (e.g., resonant or spinning disk systems), and tailored environmental chambers.

Beyond dynamic imaging, ISO-STED offers distinct advantages in thick or scattering tissues. Unlike 4Pi-based approaches that rely on delicate interference between counter-propagating beams, ISO-STED synthesizes the depletion field directly, making it inherently robust to refractive index mismatches and sample-induced aberrations [15]. We demonstrate isotropic super-resolution imaging at depths up to 30 μm in mouse brain and cartilage tissue—depths that challenge conventional 4Pi-STED even with adaptive optics or tissue clearing [17]. This robustness highlights ISO-STED's suitability for intact tissues and complex biological environments. Future integration with adaptive optics and deep-learning–based aberration correction could further enhance depth penetration and image quality in highly heterogeneous samples [36].

In addition to its technical strengths, ISO-STED enables, to our knowledge, the first 3D super-resolution mapping of mitochondrial morphology across the zonal structure of articular cartilage. Cartilage exhibits a highly stratified organization, with distinct extracellular matrix composition, chondrocyte shape, and mechanical loading across its superficial, middle, and subchondral layers [28]. While such macro- and cellular-level zonation is well documented, nanoscale differences in organelle architecture have remained elusive. ISO-STED resolves significant differences in mitochondrial volume between zones, revealing substantially enlarged mitochondria in subchondral chondrocytes relative to those in the superficial and middle layers. These findings suggest that mitochondrial scaling may reflect regional metabolic demands or biomechanical adaptations, providing new insights into how subcellular architecture reflects tissue function and potentially into early mechanisms of cartilage degeneration.

Altogether, ISO-STED emerges as an important development in the journey of STED nanoscopy. It presents a practical, cost-conscious, and accessible way to achieve isotropic 3D resolution, avoiding the complexities of dual-objective systems. Its compatibility with standard optics, simple beamline design, and robust performance in scattering media make it a versatile tool for a wide range of applications. Looking ahead, combining ISO-STED with adaptive optics [37] and deep-learning–based reconstruction methods could potentially improve depth performance and temporal resolution [38, 39]. Beyond fluorescence depletion, the core optical concepts of ISO-STED might also find applications in other areas, such as nonlinear excitation microscopy or high-precision photolithography, where precise control of isotropic fields is becoming more important.

## Method and Materials

### Nanoscope setup

The detailed design and schematic diagrams are provided in Supplementary Figure 2. The ISO-STED nanoscope is composed of the following key modules:

#### Illumination module

A pulsed 775 nm laser (MPB Communications, 80 MHz repetition rate, maximum output 3 W) serves as the STED beam, while a 635 nm picosecond pulsed laser (LDH-P-C-635, PicoQuant) functions as the excitation source. The power of both beams is independently adjusted using a combination of a half-wave plate and a polarizing beam splitter (PBS), controlled by laser control software.

The STED beam is expanded from a 1 mm diameter to 8 mm using a pair of relay lenses (AC254-050-A-ML and AC254-400-A-ML, Thorlabs) to match the active area of a reflective spatial light modulator (SLM; PLUTO-2.1-NIR-145, Holoeye). The beam is incident on the SLM at a 6° angle and then relayed by another lens pair (AC254-200-A-ML and AC254-400-A-ML, Thorlabs) to a 32-order vortex retarder (VR32-775, LBTEK), which is used for polarization extraction. Precise alignment of the optical conjugate plane and normal incidence on the center of the retarder is essential to avoid aberrations. The STED beam is then directed through a dichroic mirror (FF662-FDi01, Semrock) and conjugated onto the fast X-axis scanning mirror.

The excitation beam is expanded to 8 mm using another relay lens pair (AC254-030-A-ML and AC254-125-A-ML, Thorlabs), passed through a quarter-wave plate (AQWP05M-600, Thorlabs) to convert the polarization to left-handed circular, and merged with the STED beam using the aforementioned dichroic mirror. The combined beam is delivered to an XY galvanometric scanner (8310K, Cambridge Technology), relayed by a lens pair (89683, Edmund Optics; TTL200, Thorlabs), and focused through a 63×/1.40 NA oil-immersion objective (HCX Plan Apo, Leica Microsystems). Axial scanning is achieved using a piezoelectric objective scanner (P-725.xCDE2, Physik Instrumente), while coarse sample positioning and focus adjustment are performed using a motorized stage (Sutter Instrument).

**Fluorescence detection**

Fluorescence emitted from the sample is collected by the objective, descanned by the galvo mirrors, and separated from the excitation and STED beams using dichroic mirrors (FF662-FDi01 and FF735-Di01, Semrock). The fluorescence is further filtered with a bandpass filter (690/50, Chroma), coupled into a multimode fiber (M42L01, Thorlabs) that acts as a confocal pinhole, and detected by a single-photon counting module (SPCM-AQRH-13, Excelitas Technologies).

**Temporal synchronization**

Temporal synchronization between the excitation and STED pulses is achieved using the synchronization signal from the STED laser. This signal is delayed via an external delay generator and used to trigger the excitation laser, ensuring optimal temporal overlap and STED depletion efficiency.

**Light field polarization modulation module**

This module comprises the spatial light modulator (SLM) and the vortex retarder. As depicted in Supplementary Figure 2, these components are aligned in a 4f optical configuration to achieve precise optical conjugation. This alignment ensures that the phase patterns generated by the SLM can be accurately extracted into distinct polarization channels by the retarder. Maintaining accurate beam incidence at the center of the retarder is crucial to avoid phase distortion and preserve the integrity of the depletion field.

### Hardware control and image acquisition

Fluorescence image acquisition, galvo control, and synchronization between axial scanning and stage movement are all performed using the SciScan software platform (Scientifica, UK), which enables line-by-line image acquisition. The maximum field of view is determined to be 42.5 μm × 42.5 μm based on fluorescent bead imaging. The field size is adjustable by tuning the galvo voltage. During 3D imaging, the axial step size is set to 40 nm.

### Excitation and STED beam alignment

In addition to temporal synchronization, precise spatial overlap between the excitation and STED beams is crucial. Spatial alignment involves two dimensions: lateral and axial. Given the wavelength difference of over 100 nm between two the beams, gold nanoparticles are used for scattering-based alignment. Lateral alignment is achieved by fine-tuning the STED beam path using pre-merge mirrors. For axial alignment, chromatic aberration between the beams is compensated by adjusting the SLM phase to ensure that the Gaussian foci of both beams coincide. Axial misalignment not only compromises axial resolution but also reduces lateral resolution and the overall signal-to-noise ratio.

After coarse alignment using gold nanoparticles, fine-tuning is performed using fluorescent beads. The SLM phase pattern is iteratively adjusted to minimize the axial offset and achieve optimal 3D beam overlap.

## Biological sample preparation

### Cell culture

HeLa cells are obtained from the American Type Culture Collection (ATCC) and cultured in Dulbecco's Modified Eagle Medium (DMEM; C11965500CP, Gibco) supplemented with 10% fetal bovine serum (FBS; F8318, Sigma-Aldrich) and 1% penicillin–streptomycin (15140122, Gibco). Cells are maintained at 37 °C in a humidified incubator with 5% $CO_2$. For imaging experiments, cells are seeded onto 35 mm glass-bottom dishes (P35G-1.5-14-C, MatTek) and cultured for 24 h to allow full adhesion and stable growth before fixation or staining.

**Immunolabeling of tubulin, TOMM20, and Nup153 in fixed cells**

Immunofluorescence labeling is performed on fixed HeLa cells. Cells are prepared at approximately 70% confluency on the day of staining. Fixation is carried out using pre-warmed 4% paraformaldehyde (PFA, Sigma-Aldrich, P6148) for 30 minutes at room temperature (RT), followed by three washes with phosphate-buffered saline (PBS, 10010023, Gibco) for 5 minutes each. Cells are permeabilized with 0.5% Triton X-100 (9036-19-5, Sigma-Aldrich) for 30 minutes and washed again three times with PBS. Blocking is performed using 2% bovine serum albumin (BSA, 05470, Sigma-Aldrich) for 1 hour at RT. Primary antibodies are diluted in blocking buffer and incubated overnight at 4 °C: anti–β-tubulin (1:1000; ab6046, Abcam), and anti-TOMM20 (1:200; EPR15581-54, Abcam). The following day, cells are washed with PBS and incubated with goat anti-rabbit STAR RED-conjugated secondary antibody (1:200; Abberior) for 1 hour at RT. After three final PBS washes, coverslips are mounted with antifade mounting medium (ProLong™ Gold Antifade Mountant, P36930; Thermo Fisher) and stored overnight at 4 °C before imaging. Nup153-labeled samples are provided by Leica Microsystems (STED sample 775, product no. 158005615).

**Inner mitochondrial membrane labeling in live cells**

HeLa cells at approximately 70% confluency are labeled using PK Mito Deep Red (PKMDR-2; GENVIVO Tech.) to stain the inner mitochondrial membrane. A 250 µM stock solution is freshly prepared by dissolving 5 nmol of lyophilized dye in 20 µL of DMSO. The stock solution is diluted 1000–5000× in pre-warmed complete culture medium. The original culture medium is removed and replaced with the dye-containing medium, and cells are incubated at 37 °C for 15 minutes. Cells are then washed 1–2 times with pre-warmed medium and immediately imaged under live-cell conditions.

**Outer mitochondrial membrane labeling in mouse cartilage**

A C57BL/6 mouse is anesthetized and sacrificed in accordance with approved institutional guidelines. Limb cartilage tissues are harvested and post-fixed in 4% PFA (Sigma-Aldrich) in PBS for 24 hours at 4 °C, then stored at –80 °C. The samples are embedded in optimal cutting temperature (O.C.T.) compound (23-730-571, Fisher Healthcare) and sectioned at a thickness of 10 µm using a cryostat (CM1950, Leica). The sections are rinsed three times with PBS (5 minutes each), incubated in 3% hydrogen peroxide for 25 minutes to block non-specific proteins, and permeabilized with 0.3% Triton X-100 for 15 minutes. After blocking in 2% BSA for 1 hour at room temperature, the sections are incubated with anti-TOMM20 antibody (1:200, ab186735; Abcam) overnight at 4 °C. The following day, the sections are incubated with goat anti-rabbit STAR RED-conjugated secondary antibody (1:200;

Abberior) for 1 hour at room temperature, washed in PBS, and mounted on coverslips. The slides are dried overnight at 4 °C before imaging.

**Hematoxylin and eosin (H&E) staining of mouse cartilage**

Cryosections (10 µm) of mouse limb cartilage are fixed in 4% PFA (Sigma-Aldrich, P6148) for 30 seconds to 1 minute at room temperature (RT) and briefly rinsed in distilled water. Nuclei are stained with hematoxylin solution (Solarbio, G1120) for 3–5 minutes. Sections are differentiated in 1% acid alcohol (1% HCl in 70% ethanol) for 5–10 seconds and blued in 0.2% ammonia water (Sinopharm Chemical Reagent Co.) for 20 seconds. After rinsing, sections are counterstained with eosin Y solution (Solarbio, G1100) for 10–20 seconds. Dehydration is performed through an ethanol gradient (70%, 80%, 95%, and 100%, 1 min each), followed by xylene clearing (Sinopharm, analytical grade) and mounting with neutral balsam (Solarbio, G8590). Brightfield images are acquired using a Leica DM5000 microscope.

**Neuron labeling with mice brain slice**

Adult eYFP-H transgenic mice (~12 weeks old) are used for neuron labeling. Mice are deeply anesthetized with isoflurane (5% flow rate, SomnoFlo; Kent Scientific) and transcardially perfused with ice-cold 1× PBS, followed by ice-cold 4% PFA in PBS. After perfusion, brains are carefully harvested and post-fixed overnight at 4 °C in 4% PFA. The following day, brains are washed three times in 1× PBS (15 minutes each wash) before further processing.

For sectioning, brains are embedded in low-melt agarose and coronally sliced at a thickness of 75 µm using a vibratome (Leica VT1200S). The sections are collected into PBS and subjected to immunostaining, following an adapted iDISCO protocol (https://www.cell.com/fulltext/S0092-8674%2814%2901297-5) optimized for thicker tissue slices to ensure complete antibody penetration. All staining steps are performed in 500 mL amber tubes, with samples continuously agitated at 50 rpm either at room temperature (RT) or 37 °C as appropriate.

Tissue sections are first washed twice for 1 hour each in PBS containing 0.2% Triton X-100 (PTx.2) at room temperature (RT) to remove residual fixative. Samples are then incubated overnight at 37 °C in PBS containing 0.2% Triton X-100 and 20% dimethyl sulfoxide (DMSO), followed by another overnight incubation at 37 °C in PBS supplemented with 0.1% Tween-20, 0.1% Triton X-100, 0.1% deoxycholate, 0.1% NP-40, and 20% DMSO. After pretreatment, sections are washed twice again in PTx.2 at RT.

For immunolabelling, slices are first incubated in permeabilization solution at 37 °C for 4–6 hours, followed by blocking in blocking solution at 37 °C for an additional 4–6 hours. Samples are then incubated with the primary antibody (rabbit anti-GFP, 1:500; NB600-308, Bio-Techne) diluted in PTwH buffer (PBS containing 0.1% Tween-20 and 10 µg/mL heparin) supplemented with 5% DMSO and 3% goat serum at 37 °C overnight. After extensive washing in PTwH at room temperature (RT) (four to five times, 15–20 minutes per wash), sections are incubated with a far-red conjugated secondary antibody (1:200; STAR RED, Abberior) diluted in PTwH containing 3% goat serum at 37 °C overnight. Samples are then washed again extensively in PTwH at RT. Following immunolabeling, slices are floated onto glass slides in PBS, allowed to partially dry at RT, and mounted with antifade mounting medium under 1.5 mm thick coverslips. Mounted samples are stored at 4 °C until imaging.

## Image processing and visualization

Image processing and quantitative analysis are performed using GraphPad Prism 10 and MATLAB R2019b (MathWorks). Deconvolution of all images is carried out using either Huygens Professional (Scientific Volume Imaging) or the Richardson–Lucy deconvolution algorithm implemented in the DeconvolutionLab2 plugin for FIJI (ImageJ). The Z-stack Depth Color Code plugin in FIJI is used to pseudo-color 3D images according to depth. Mitochondrial volume measurements are performed using the 3D Objects Counter plugin in FIJI. Imaris (Oxford Instruments) is used for 3D rendering and visualization. Videos are generated using Adobe Premiere Pro 2021 (Adobe Inc.).

## Data Availability

All data supporting the findings of this study are available from the corresponding author on request.

# References

1. Sahl S. J., Hell S. W., Jakobs S. Fluorescence nanoscopy in cell biology. *Nat. Rev. Mol. Cell Biol.* **18**, 685-701 (2017).

2. Balasubramanian H., Hobson C. M., Chew T.-L., et al. Imagining the future of optical microscopy: everything, everywhere, all at once. *Commun. Biol.* **6**, 1096 (2023).

3. Alvelid J., Damenti M., Sgattoni C., et al. Event-triggered STED imaging. *Nat. Methods* **19**, 1268-1275 (2022).

4. Mudry, E., Le, M. E., Ferrand, P., Chaumet, P. C., Sentenac, A. Isotropic diffraction-limited focusing using a single objective lens. *Phys. Rev. Lett.* **105**, 203903 (2010).

5. Hell S., Stelzer E. H. K. Properties of a 4Pi confocal fluorescence microscope. *J. Opt. Soc. Am. A* **9**, 2159-2166 (1992).

6. Hell S. W. Far-Field Optical Nanoscopy. Science **316**, 1153-1158 (2007).

7. Shao L., Kner P., Rego E. H., et al. Super-resolution 3D microscopy of live whole cells using structured illumination. *Nat. Methods* **8**, 1044-1046 (2011).

8. Rust M. J., Bates M., Zhuang X. Sub-diffraction-limit imaging by stochastic optical reconstruction microscopy (STORM). *Nat. Methods* **3**, 793-796 (2006).

9. Lukinavičius G., Alvelid J., Gerasimaitė R., et al. Stimulated emission depletion microscopy. Nat. Rev. Methods Primers **4**, 56 (2024).

10. Betzig E., Patterson G. H., Sougrat R., et al. Imaging Intracellular Fluorescent Proteins at Nanometer Resolution. Science **313**, 1642-1645(2006).

11. Huang B.,Wang W., Bates M., et al. Three-Dimensional Super-Resolution Imaging by Stochastic Optical Reconstruction Microscopy. Science **319**, 810-813 (2008).

12. Wildanger D., Medda R., Kastrup L., et al. A compact STED microscope providing 3D nanoscale resolution. *J. Microsc.* **236**, 35-43 (2009).

13. Harke B., Ullal C. K., Keller J., et al. Three-Dimensional Nanoscopy of Colloidal Crystals. *Nano Lett.* **8**, 1309-1313 (2008).

14. Hell S. W., Nagorni M. 4Pi confocal microscopy with alternate interference. *Opt. Lett.* **23**, 1567-1569 (1998).

15. Ouyang Z., Wang Q., Li X., et al. Elucidating subcellular architecture and dynamics at isotropic 100-nm resolution with 4Pi-SIM. *Nat. Methods* **22**, 335-347 (2025).

16. Schmidt R., Wurm C. A., Jakobs S., et al. Spherical nanosized focal spot unravels the interior of cells. *Nat. Methods* **5**, 539-544 (2008).

17. Hao X., Allgeyer E. S., Lee D.-R., et al. Three-dimensional adaptive optical nanoscopy for thick specimen imaging at sub-50-nm resolution. *Nat. Methods* **18**, 688-693 (2021).

18. Bates M., Keller-Findeisen J., Przybylski A., et al. Optimal precision and accuracy in 4Pi-STORM using dynamic spline PSF models. *Nat. Methods* **19**, 603-612 (2022).

19. Shtengel G., Galbraith J. A., Galbraith C. G., et al. Interferometric fluorescent super-resolution microscopy resolves 3D cellular ultrastructure. *Proc. Natl. Acad. Sci.* **106**, 3125-3130 (2009).

20. Böhm U., Hell S. W., Schmidt R. 4Pi-RESOLFT nanoscopy. *Nat. Commun.* **7**, 10504 (2016).

21. Huang F., Sirinakis G., Allgeyer E. S., et al. Ultra-High Resolution 3D Imaging of Whole Cells. Cell. **166**, 1028-1040 (2016).

22. Weng X., Song Q., Li X., et al. Free-space creation of ultralong anti-diffracting beam with multiple energy oscillations adjusted using optical pen. *Nat. Commun.* **9**, 5035 (2018).

23. Weng X., Miao Y., Zhang Q., et al. Extraction of Inherent Polarization Modes from an m‑Order Vector Vortex Beam. *Adv. Photon. Res.* **3**, 10 (2022).

24. Milione G., Lavery M. P. J., Huang H., et al. 4 × 20 Gbit/s mode division multiplexing over free space using vector modes and a q-plate mode (de)multiplexer. *Opt. Lett.* **40**, 1980-1983 (2015).

25. Moreno I., Sanchez-Lopez M. M., Badham K., et al. Generation of integer and fractional vector beams with q-plates encoded onto a spatial light modulator. *Opt. Lett.* **41**, 1305-1308 (2016).

26. Prashar A., Bussi C., Fearns A., et al. Lysosomes drive the piecemeal removal of mitochondrial inner membrane. *Nature* **632**, 1110-1117 (2024)

27. Winstanley Y. E., Rose R. D., Sobinoff A. P., et al. Telomere length in offspring is determined by mitochondrial-nuclear communication at fertilization. *Nat. Commun.* **16**, 2527 (2025).

28. Pueyo Moliner A., Ito K., Zaucke F., et al. Restoring articular cartilage: insights from structure, composition and development. *Nat. Rev. Rheumatol.*, **21**, 291-308 (2025).

29. Peters J. R., Hoogenboom M., Abinzano F., et al. Tissue growth as a mechanism for collagen fiber alignment in articular cartilage. *Sci. Rep.* **14**, 31121 (2024).

30. Danalache M., Beutler K. R., Rolauffs B., et al. Exploration of changes in spatial chondrocyte organisation in human osteoarthritic cartilage by means of 3D imaging. *Sci. Rep.* **11**, 9783 (2021).

31. Hu Y., Chen X., Wang S., et al. Subchondral bone microenvironment in osteoarthritis and pain. *Bone Res.* **9**, 20 (2021).

32. Liu D., Cai Z. J., Yang Y. T., et al. Mitochondrial quality control in cartilage damage and osteoarthritis: new insights and potential therapeutic targets. *Osteoarthr. Cartil.* **30**, 395-405 (2022).

33. Hell S. W., Wichmann J. Breaking the diffraction resolution limit by stimulated emission: stimulated-emission-depletion fluorescence microscopy. *Opt Lett.* **19**, 780-782 (1994).

34. Osseforth C., Moffitt J. R., Schermelleh L., et al. Simultaneous dual-color 3D STED microscopy. *Opt. Express* **22**, 7028-7039 (2014).

35. Yang X., Yang Z., Wu Z., et al. Mitochondrial dynamics quantitatively revealed by STED nanoscopy with an enhanced squaraine variant probe. *Nat. Commun.* **11**, 3699 (2020).

36. Guo M., Wu Y., Hobson C. M., et al. Deep learning-based aberration compensation improves contrast and resolution in fluorescence microscopy. *Nat. Commun.* **16**, 313 (2025).

37. Tu S., Liu X., Yuan D., et al. Accurate Background Reduction in Adaptive Optical Three-Dimensional Stimulated Emission Depletion Nanoscopy by Dynamic Phase Switching. *ACS Photonics* **9**, 3863-3868 (2022).

38. Fu S., Shi W., Luo T., et al. Field-dependent deep learning enables high-throughput whole-cell 3D super-resolution imaging. *Nat. Methods* **20**, 459-468 (2023).

39. Qiao C., Li D., Guo Y., et al. Evaluation and development of deep neural networks for image super-resolution in optical microscopy. *Nat. Methods* **18**, 194-202 (2021).

**Acknowledgments**

Parts of this work were supported by the National Key R&D Program of China (2021YFF0502900), National Natural Science Foundation of China (T2421003/62375183/62127819/62435011), Shenzhen Key Laboratory of Photonics and Biophotonics (ZDSYS20210623092006020), Shenzhen Science and Technology Program (JCYJ20220818100202005).

**Author contributions**

R.Z. and W.X. designed the microscope system, performed data analysis, and wrote the manuscript. R.Z. and H.Z. constructed the imaging setup. X.W. and J.Q. conceived the study and experimental design. X.W. and R.Z. carried out numerical simulations. L.W., F.L., W.Y., D.L., and C.Z. performed experimental validation. K.K.G. and E.R.E.S. prepared the mouse brain tissue samples. X.G., B.Y., and L.L. contributed to manuscript writing, review, and revision. S.Z. offered advice regarding its development. J.Q. directed the entire project.

**Competing interests statement**

The authors declare that they have no competing financial and non-financial interests to disclose.

# Figures

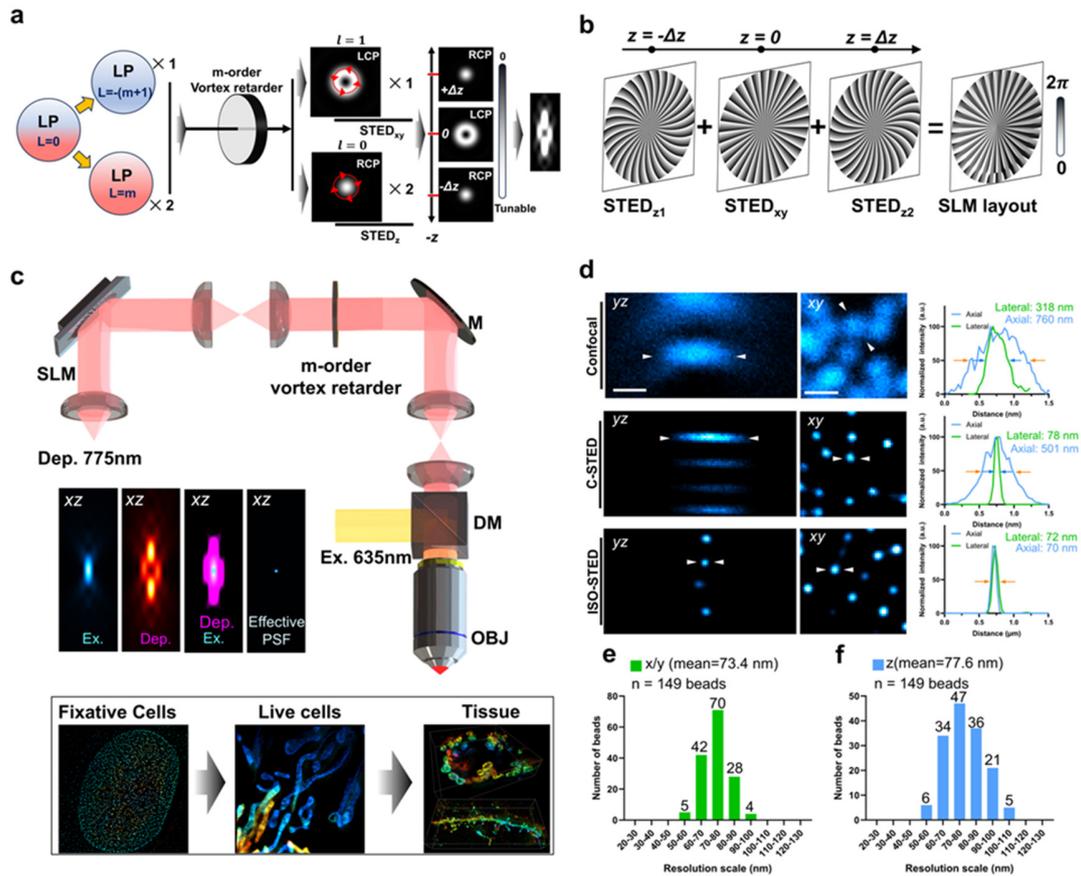

**Figure 1** Principle and implementation of single-objective, single-depletion-beam isotropic 3D STED (ISO-STED) microscopy. a, Schematic of depletion beam generation. A linearly polarized beam with zero topological charge is modulated by an optical pen to produce one beam with a topological charge of –(m+1) and two beams with a topological charge of +m. These three beams are then passed through an m-order vortex retarder to extract one left-circularly polarized beam carrying a topological charge of +1 and two right-circularly polarized beams with zero topological charge. By precisely tuning the axial positions and intensity distributions of these foci, a tailored depletion beam is formed to achieve isotropic 3D super-resolution. b, Principle of phase generation. Focal points positioned at z = −Δz and z = +Δz are assigned orbital angular momentum (OAM) of +m, while the central focal point (z = 0) is assigned OAM of –(m+1). The relative intensities of these beams are controlled via the optical pen, and their combined phase pattern is encoded onto the SLM. c, Optical layout. A 775 nm nanosecond depletion beam is modulated by the SLM and relayed via a pair of 4f systems to the m-order vortex retarder, then conjugated to the back aperture of the objective lens. Simultaneously, a 635 nm excitation beam is spatiotemporally aligned and co-focused onto the sample. Simulated intensity distributions of the excitation beam, depletion beam, and their merged configuration are shown, along with the resulting effective point spread function (PSF). The system supports multiscale, isotropic 3D super-resolution imaging in fixed cells, live cells, and thick tissues. d, Representative imaging of 20 nm fluorescent beads using confocal, conventional 2D-STED, and ISO-STED in both xy and xz views. The measured full width at half maximum (FWHM) values in the lateral direction are 318 nm (confocal), 78 nm (2D-STED), and 72 nm (ISO-STED); in the axial direction, they are 760 nm (confocal), 501 nm (2D-STED), and 70 nm (ISO-STED), scale bar = 500nm. e, Statistical distribution of lateral resolution measured from 149 fluorescent beads using ISO-STED. f, Statistical distribution of axial resolution measured from the same set of beads using ISO-STED, the mean value of the lateral and axial resolution are 73.4 nm and 77.6 nm, respectively.

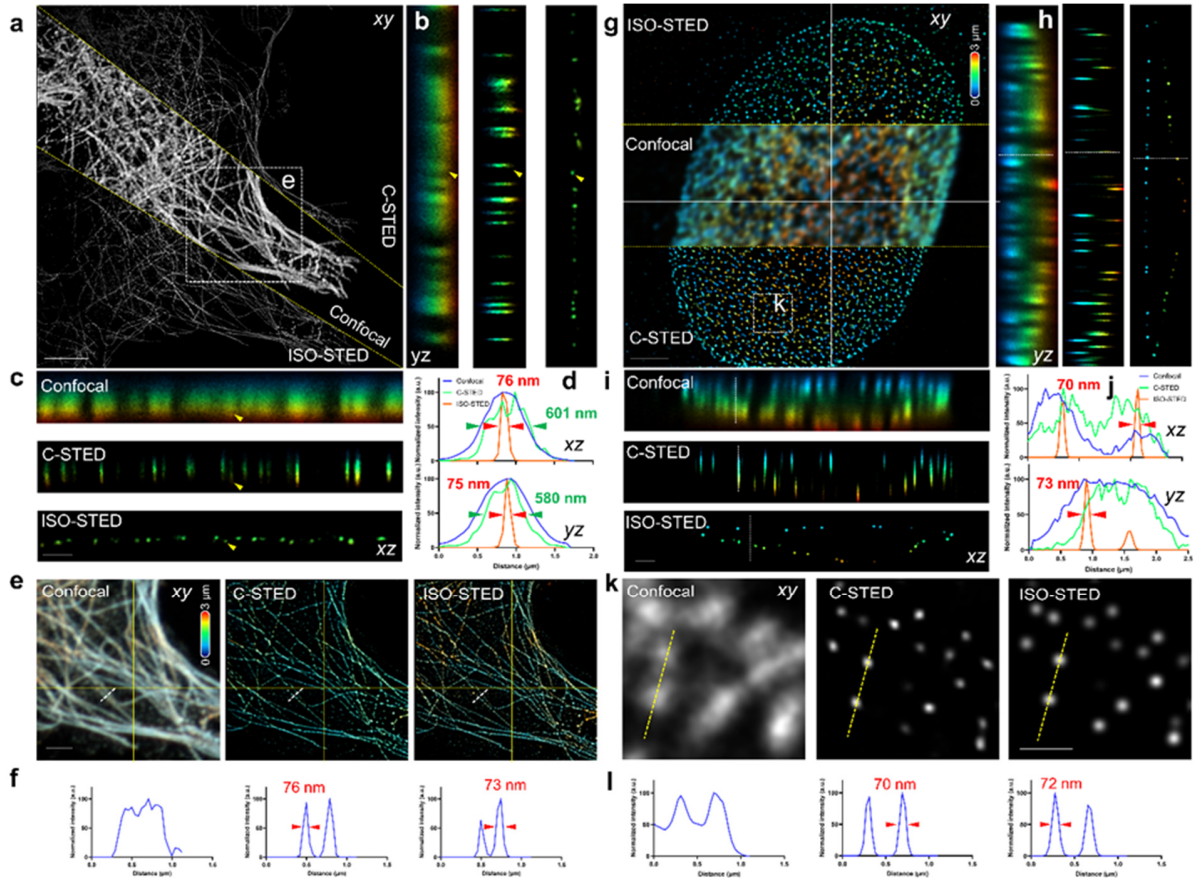

**Figure 2** ISO-STED enables isotropic 3D imaging of fixed cells. a, Confocal, conventional STED (C-STED), and ISO-STED imaging of tubulin in the x-y plane. Scale bar = 5 μm. b, c, xz and yz cross-sections along the yellow line in e for each imaging modality. Scale bar = 1 μm. d, Upper panels: Intensity profiles along the z-axis at the positions indicated by arrows in the x-z-cross-section (b), with FWHM values of 76 nm (ISO-STED) and 601 nm (confocal). Lower panels: Intensity profiles along the z-axis at the positions indicated by arrows in the y-z-cross-section (c), with FWHM values of 75 nm (ISO-STED) and 580 nm (confocal). e, Magnified view from the white box in a, showing the projection of the three modalities along the imaging depth range of z = 0–3 μm, with depth encoded by the color bar. Scale bar = 2 μm. f, Intensity distribution along the white dashed line in e, showing the FWHM of the peaks in conventional STED and ISO-STED as 76 nm and 73 nm, respectively. g, Confocal, C-STED, and ISO-STED imaging of nup153 along the z = 0–3 μm imaging depth range, with depth encoded by the color bar. Scale bar = 5 μm. h, i, x-z and y-z cross-sections along the white line in g for each imaging modality. Scale bar = 1 μm. j, Upper panels: Intensity profiles along the white dashed line in h for each modality, with FWHM values of 70 nm (ISO-STED). Lower panels: Intensity profiles along the white line in the y-z-cross-section (i) for each modality, with FWHM values of 73 nm (ISO-STED). k, Magnified xy-view at z = 0.85 μm from the boxed region in g. Scale bar = 500 nm. l, Intensity distribution along the yellow dashed line in k, with FWHM values of 70 nm (conventional STED) and 72 nm (ISO-STED). Scale bars: 5 μm (c, f); 2 μm (a, projection of k, l); 1 μm (b, d); and 500 nm (g, h, m). The thickness of all cross-section slices is 50 nm. Representative results from three to five independent experiments are shown.

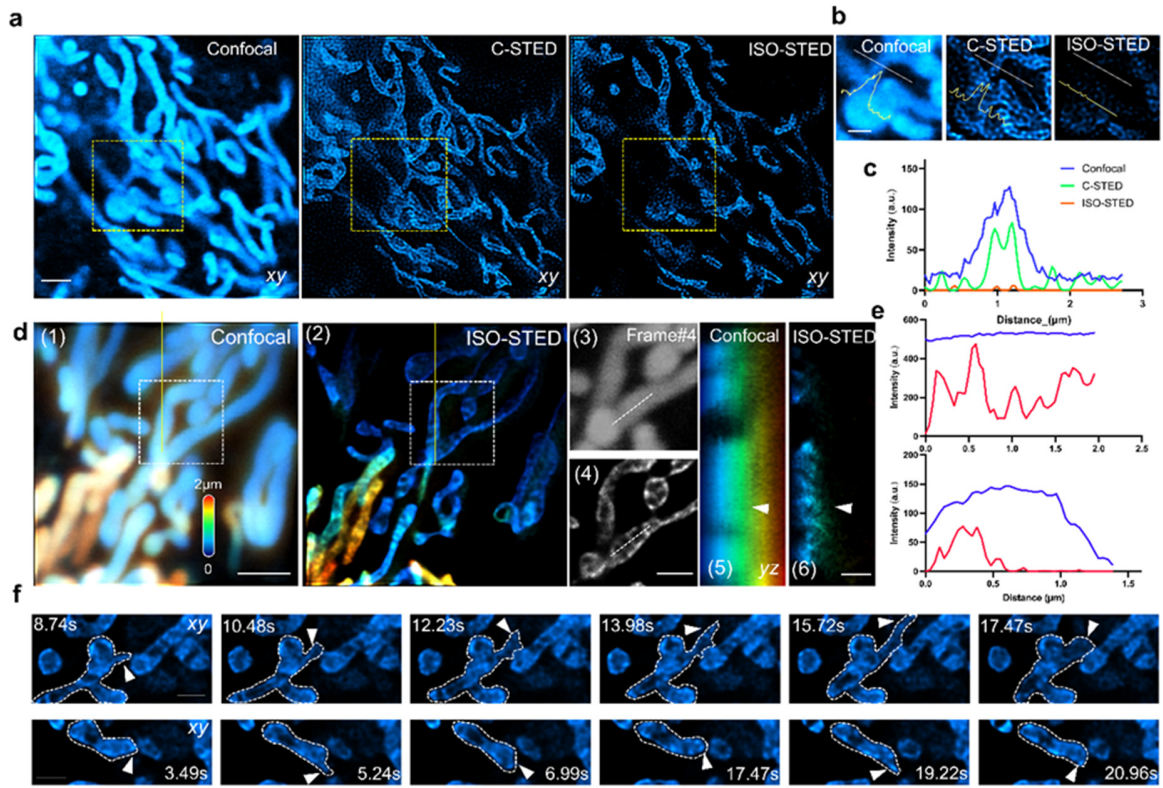

**Figure 3** ISO-STED enables high-contrast monitoring of mitochondrial dynamics by removing out-of-focus signals. a, X–Y views of Confocal, C-STED, and ISO-STED images of mitochondria. Scale bar = 5 μm. b, Magnified view of the yellow boxed region in a. c, Intensity profile along the yellow dashed line in the right section of a. d, (1) and (2) 3D reconstructions of Confocal and ISO-STED images within the z = 0–3 μm range, with depth encoded according to the color bar. Scale bar = 2 μm. (3) and (4) X–Y views of Confocal and ISO-STED at frame #4. Scale bar = 1 μm. (5) and (6) Y–Z cross-sections along the yellow solid line in (1) and (2). Scale bar = 1 μm. e, Upper panel: Intensity distribution along the white dashed line in (3) and (4). Lower panel: Intensity profiles along the z-axis at the positions indicated by the arrows in (5) and (6). f, temporal results from two different views of mitochondrial morphology, with arrows highlighting transient morphological changes in mitochondria. Scale bar = 1 μm.

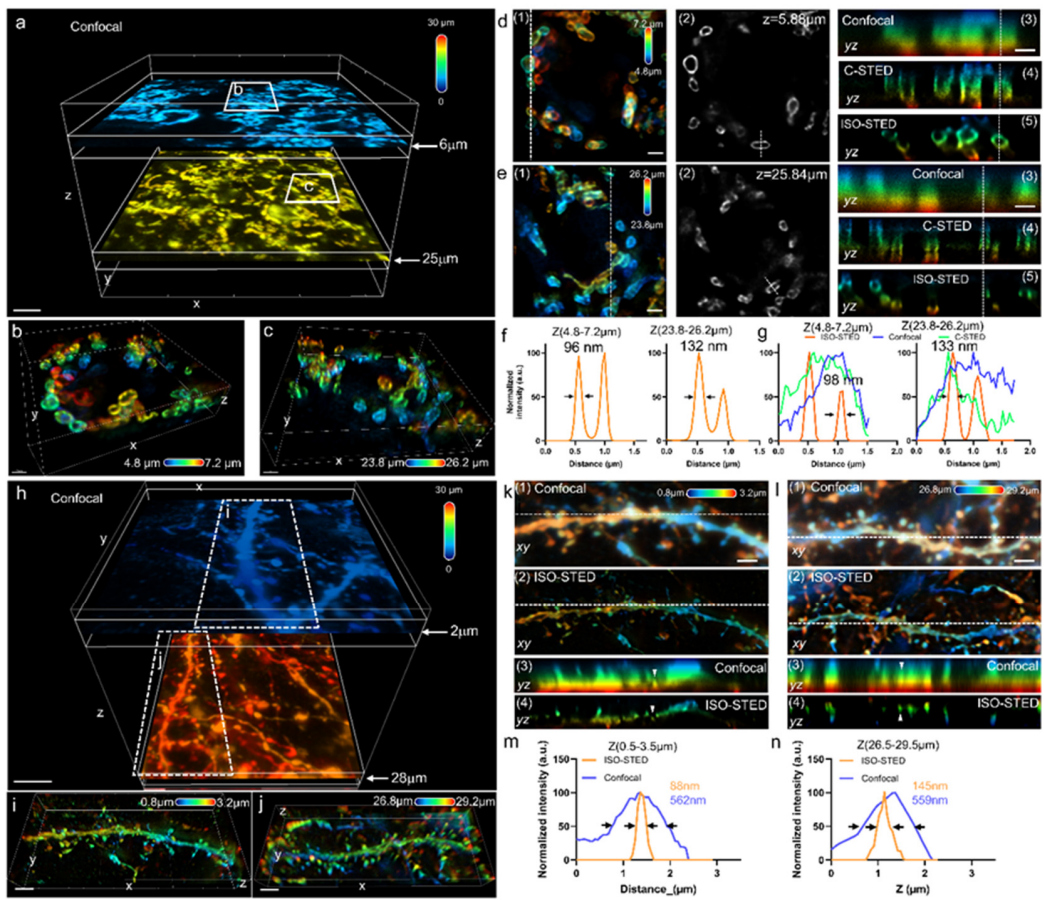

**Figure 4 ISO-STED enables isotropic super-resolution imaging in thick tissue.** a, Large-volume confocal imaging of mouse joint tissue sections labeled with a mitochondrial outer membrane marker. Imaging was performed from $z = 0$ to 30 µm, with depth color-coded. Regions shown in $b$ and $c$ correspond to sub-volumes at $z = 4.8$–7.2 µm and $z = 23.8$–26.2 µm, respectively. Scale bar = 3 µm. b, 3D reconstruction of the boxed region in $a$ at $z = 4.8$–7.2 µm, imaged using ISO-STED. Scale bar = 1 µm. c, 3D reconstruction of the boxed region in $a$ at $z = 23.8$–26.2 µm, acquired with ISO-STED. Scale bar = 1 µm. d, (1) 2D maximum intensity projection of the volume shown in $b$. Scale bar = 1 µm. (2) x–y slice at $z = 5.88$ µm. (3–5) y–z cross-sections at the dashed line position in (1), acquired with confocal, C-STED, and ISO-STED, respectively. Scale bar = 1 µm. e, (1) 2D projection of the volume in $c$. Scale bar = 1 µm. (2) x–y slice at $z = 25.84$ µm. (3–5) y–z cross-sections at the dashed line position in (1), for confocal, C-STED, and ISO-STED, respectively. Scale bar = 1 µm. f, Intensity profiles along the white dashed lines in (d2) and (e2), showing lateral resolution comparison. g, Intensity profiles along the white dashed lines in the y–z cross-sections of (d3–5) and (e3–5), showing axial resolution comparison across modalities. h, Large-volume confocal imaging of mouse brain slices labeled for neuronal structures, spanning $z = 0$ to 30 µm. Color encodes imaging depth. Regions shown in $i$ and $j$ correspond to sub-volumes at $z = 0.8$–3.2 µm and $z = 26.8$–29.2 µm, respectively. Scale bar = 3 µm. i, 3D reconstruction of the boxed region in $h$ at $z = 0.8$–3.2 µm using ISO-STED. Scale bar = 3 µm. j, 3D reconstruction of the boxed region in $h$ at $z = 26.8$–29.2 µm using ISO-STED. Scale bar = 3 µm. k, (1) x–y projection of the region in $i$ imaged with confocal. Scale bar = 2 µm. (2) x–y projection of the same region imaged with ISO-STED. (3, 4) y–z cross-sections at the dashed line positions in (1) and (2) for confocal and ISO-STED, respectively. l, (1) x–y projection of the region in $j$ imaged with confocal. Scale bar = 2 µm. (2) ISO-STED x–y projection of the same region. (3, 4) y–z cross-sections at the dashed line positions in (1) and (2) for confocal and ISO-STED, respectively. m, Intensity profiles along the z-direction at the positions indicated by white arrows in (k3) and (k4), comparing confocal and ISO-STED. n, Intensity profiles along the z-direction at the positions indicated by white arrows in (l3) and (l4), comparing confocal and ISO-STED.z

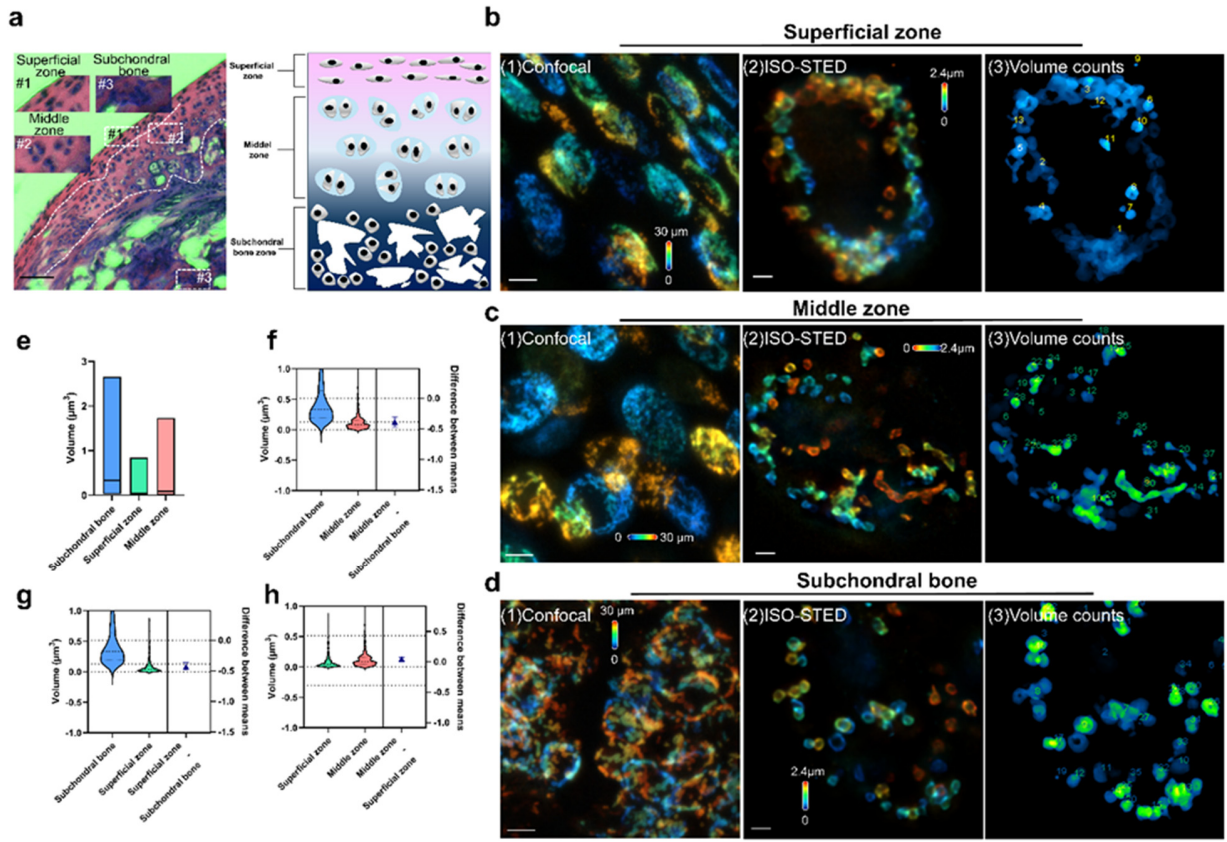

**Figure 5** ISO-STED reveals mitochondrial volume differences across cartilage layers. a, Low-magnification H&E-stained section of mouse articular cartilage, showing three distinct zones—superficial, middle, and subchondral bone—distinguished by differences in cell morphology. A schematic on the right illustrates the zonal organization. Scale bar = 50 μm. b–d, Representative imaging results from the superficial zone (b), middle zone (c), and subchondral bone zone (d). (1) Confocal X–Y projections acquired with a 300 nm Z-step across a 30 μm depth. Color encodes axial depth. Scale bar = 5 μm. (2) ISO-STED x–y projections from the top 2.4 μm imaging range, acquired with a 40 nm z-step. Scale bar = 2 μm. (3) Mitochondria identified and segmented using the 3D Objects Counter plugin in FIJI, with individual objects rendered in false color. e, Quantification of mitochondrial volume across cartilage zones. Average volumes were 0.5142 μm³ in the subchondral bone zone, 0.13 μm³ in the middle zone, and 0.084 μm³ in the superficial zone. Standard deviations were 0.5028, 0.1696, and 0.1129, respectively. f–h, Statistical comparisons of mitochondrial volumes between cartilage zones using unpaired two-tailed t tests: f, subchondral bone vs. middle zone, P < 0.0001; g, subchondral bone vs. superficial zone, P < 0.0001; h, middle zone vs. superficial zone, P = 0.0013.

# Supplementary Information

## Three-Dimensional Isotropic STED Nanoscopy using a Single Objective

Zhang et al.

**TABLE OF CONTENTS**

**Supplementary Notes**

**Supplementary Figures**

**Supplementary Table**

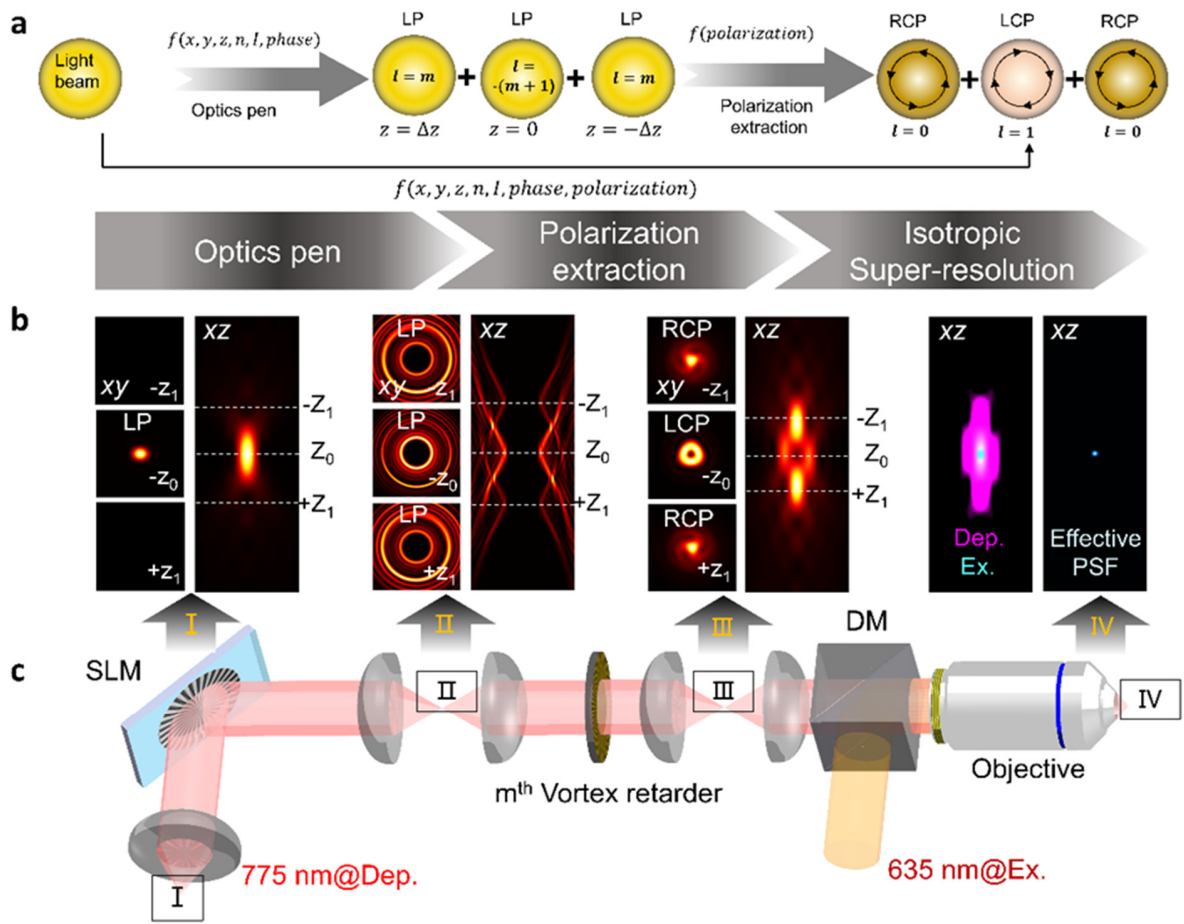

**Supplementary Figure 1** Principle of isotropic depletion foci generation in ISO-STED. a, Schematic workflow illustrating the generation of the ISO-STED depletion point spread function (PSF). Using the optical pen strategy, three foci are engineered with independent control over spatial position ($x$, $y$, $z$), foci number, energy, and phase (including orbital angular momentum, OAM). These foci are axially spaced at $z = +\Delta z$, 0, and $-\Delta z$, and are assigned OAM values of $m$, $-(m+1)$, and $m$, respectively. Initially linearly polarized, the beams are then subjected to polarization mode extraction: the two outer foci (STED$_z$) are converted to right-circular polarization (RCP), while the central focus (STED$_{xy}$) is converted to left-circular polarization (LCP). From left to right, the flow chart illustrates comprehensive modulation of spatial location, phase, energy, and polarization state, resulting in the generation of an isotropic depletion field. b–c, Sequential stages (I–IV) correspond to the labeled panels in c showing the focal intensity distributions in $x$-$y$ and $x$-$z$ dimension. I, A single Gaussian focus with no phase modulation. II, three distinct OAM-modulated foci (topological charges of $+m$, $-(m+1)$, and $+m$ are positioned at $z = +\Delta z$, 0, and $-\Delta z$. III, after passing through an $m^{th}$-order vortex retarder, polarization states of the three foci are separated and re-assigned. IV, the modulated depletion beam is spatially overlapped with the excitation beam to induce stimulated emission depletion, producing a compressed, isotropic 3D effective PSF.

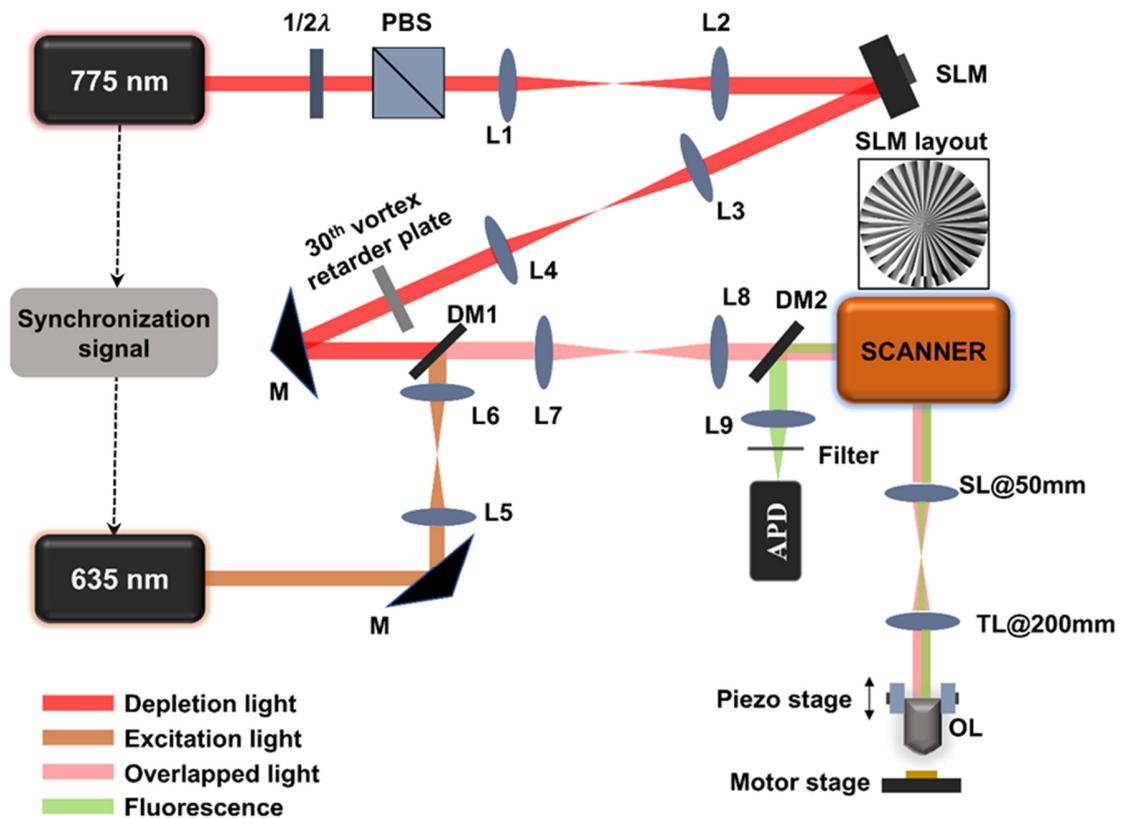

**Supplementary Figure 2** Optical layout of the ISO-STED nanoscope. The overall system architecture closely follows that of conventional 2D-STED microscopy. In this setup, a 775-nm nanosecond pulsed laser is used as the depletion beam (red path), which also provides a synchronization signal to trigger the 635-nm picosecond excitation laser (brown path), enabling precise temporal alignment of the two pulse trains. The depletion beam first passes through a combination of a half-wave plate and a polarizing beam splitter (PBS) for power adjustment and polarization control, ensuring horizontal linear polarization for optimal modulation efficiency at the spatial light modulator (SLM). After phase modulation by the SLM, the beam is relayed via a 4f system and conjugated onto a 30th-order vortex retarder plate, then further relayed by another 4f system onto the X-axis galvanometric mirror. The 635-nm excitation laser is expanded by a 4f system and spatially combined with the depletion beam at dichroic mirror 1 (pink path). The combined beams are relayed through the same 4f optical train onto the galvanometric scanning mirrors (X and Y axes), then directed through a scan lens and tube lens to fill the back aperture of a high-NA objective. For 3D imaging, axial scanning is achieved using a piezo objective scanner. Fluorescence emitted from the sample is collected by the same objective and descanned through the identical optical path. The emission is separated from the excitation and depletion light via dichroic mirror 2, filtered to remove background, and finally detected by an avalanche photodiode (APD).

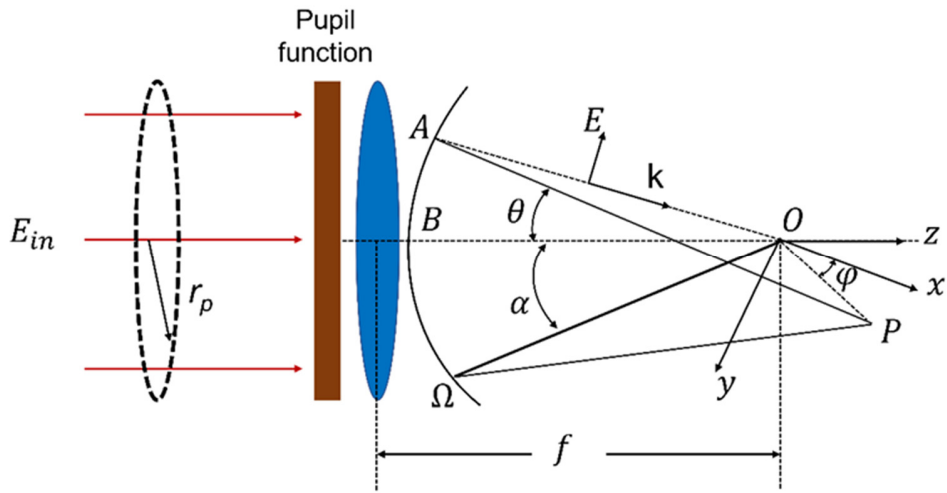

**Supplementary Figure 3** Schematic of deriving optical pen and the focusing sytem. $\Omega$ is the focal sphere, with the center at O, and a radius $f$. A, B are the off- and on-axis points in $\Omega$. $P$ is an arbitrary point in the focal region. There is a Pupil filter in the wavefront of the objective lens, $\theta$ and $\varphi$ are the convergent and azimuthal angle, respectively.

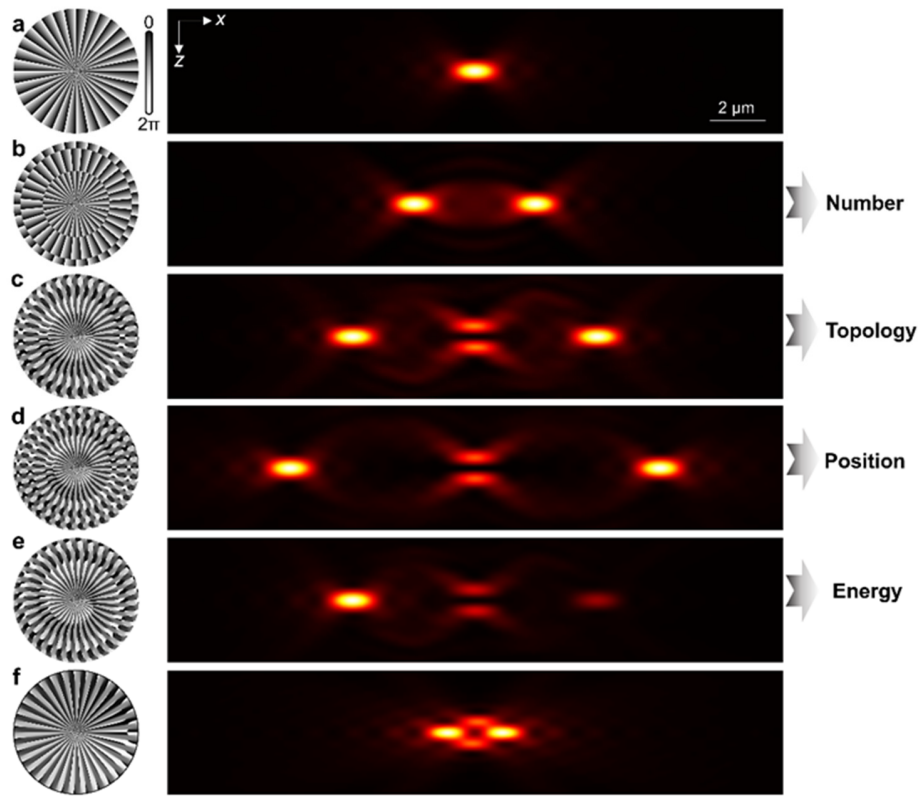

**Supplementary Figure 4** Independent modulation of foci number, phase, position, and energy using the optical pen. All panels show the corresponding phase patterns displayed on the SLM, (left) and the resulting x–z plane light field distributions (right). Scale bar, 2 μm. a, only $T = \exp(im\varphi)$ displayed on SLM, ($m$=30); a single Gaussian focus is generated. b, Generation of two axially separated foci. c, Addition of a third focus with independently tunable orbital angular momentum. d, Precise modulation of the spatial position of each focus. e, Control of the relative energy distribution among foci. f, Leveraging the modulation capabilities of these dimensions, we can ultimately generate the ISO-STED depletion PSF. All the phase parameters setting are shown in Supplementary Note. 3.

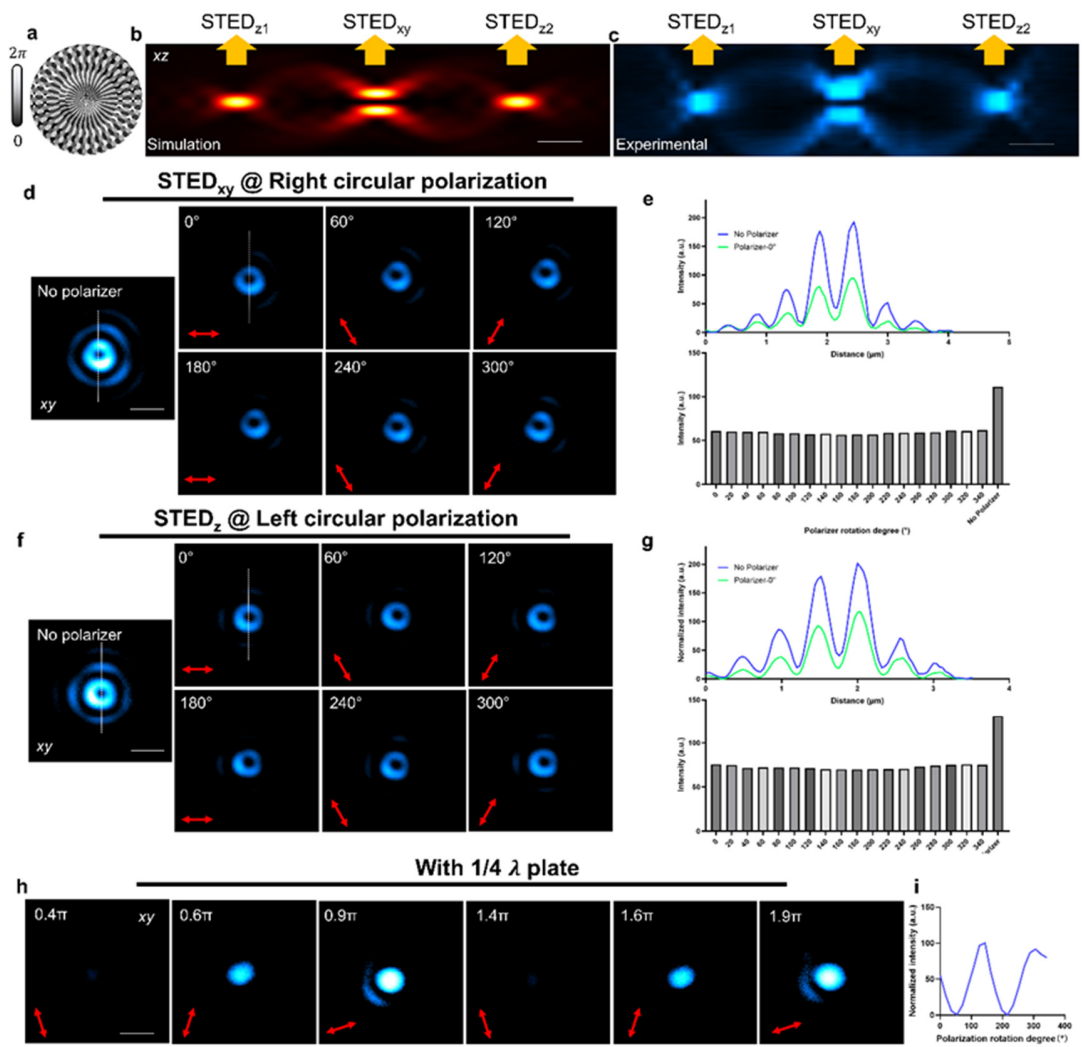

**Supplementary Figure 5** Polarization assignment and verification of ISO-STED depletion light. a, Phase pattern displayed on the SLM. All the phase parameters setting are shown in Supplementary Note. 3. b, Simulated $x$–$z$ light field distribution based on the phase in a. Scale bar, 2 μm. c, Experimental validation of the $x$–$z$ light field distribution. Scale bar, 2 μm. d, Experimental verification of RCP for the $STED_{xy}$ focus. The leftmost image shows the light field without an analyzer. The six images to the right show the light field recorded at 60° rotational increments of the analyzer. e, Top: intensity profiles along the dashed lines in the 0° and analyzer-free images in d. Bottom: quantified peak intensities of the central lobe as a function of analyzer rotation angle, showing approximately a twofold reduction in signal intensity when the analyzer is applied. f, Experimental verification of left-circular polarization for the $STED_z$ focus, following the same procedure as in d. g, Intensity analysis for f, analogous to the plots shown in e. h, Polarization analysis using a quarter-wave plate placed before the analyzer, with its fast axis oriented at 45° relative to the horizontal. Light fields are shown for analyzer rotation angles corresponding to phase delays of 0.4π, 0.6π, 0.9π, 1.4π, 1.6π, and 1.9π. i, the measured intensity as a function of analyzer angle follows a sinusoidal dependence, consistent with expected behavior for circular polarization states.

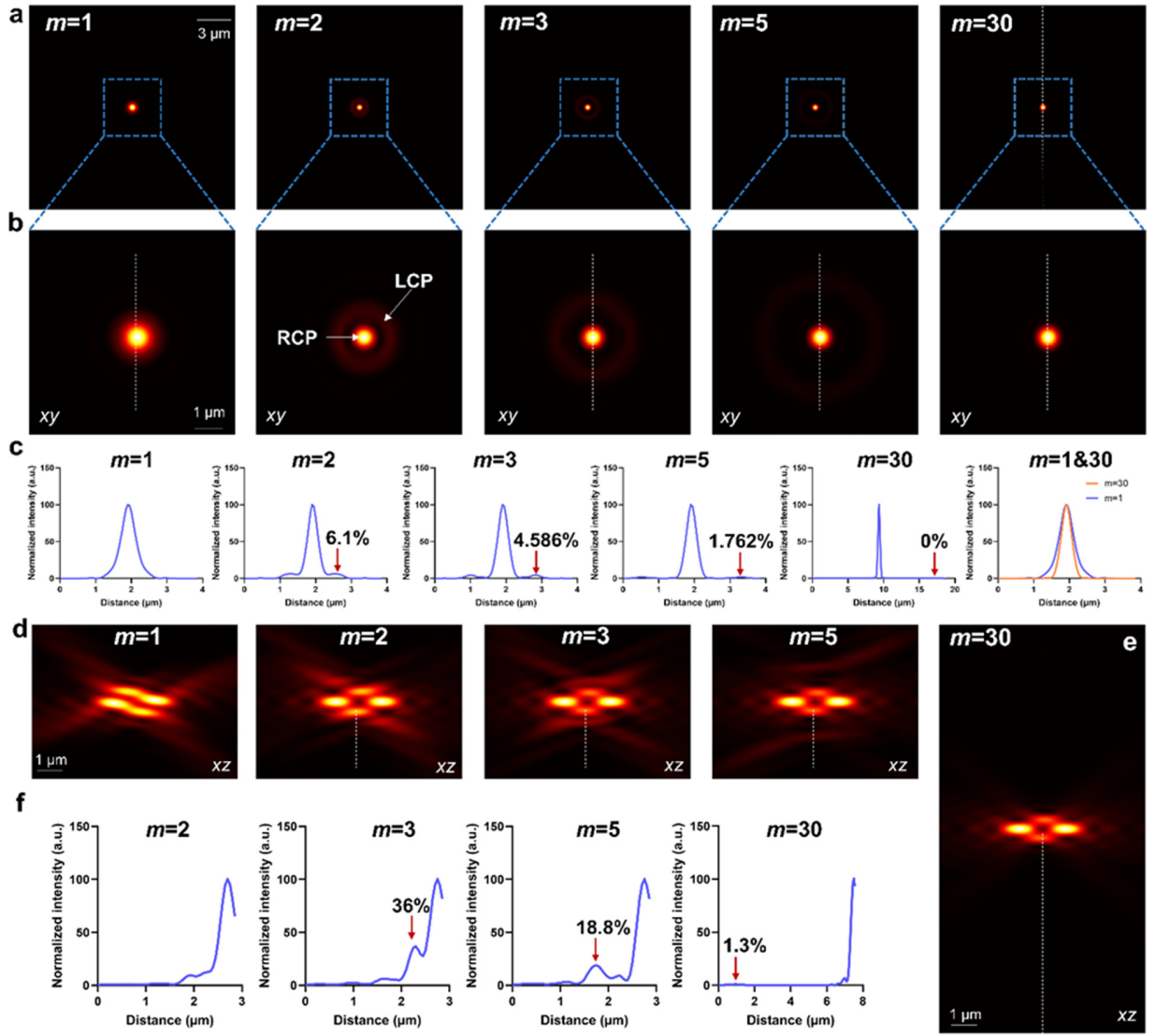

**Supplementary Figure 6** Influence of topological charge on polarization mode extraction efficiency. a, Simulated depletion field intensity distributions for vortex retarders with varying topological charges ($m$ = 1, 2, 3, 5, and 30). The SLM displays a combined phase pattern used to generate three foci. The central focus ($l = 0$) corresponds to LCP mode, while the outer ring lobes ($l = 2m$) correspond to RCP mode. Scale bar, 3 μm. b, Magnified view of the blue-dashed region in a, highlighting spatial separation of LCP and RCP modes. Scale bar, 3 μm. c, Intensity profiles along the white dashed line in b, showing side lobes corresponding to the RCP component. The final panel compares the intensity profiles for $m$ = 1 and $m$ = 30, demonstrating that $m$ = 1 fails to spatially separate polarization modes, whereas increasing $m$ enables efficient separation. Specifically, the residual RCP intensity in the STED$_{xy}$ region is reduced to 6.1% ($m$ = 2), 4.586% ($m$ = 3), 1.762% ($m$ = 5), and ~0% ($m$ = 30). d, Simulated depletion field intensity distributions ($xz$ view) for $m$ = 1, 2, 3, and 5. e, Simulated $xz$ distribution for m = 30, showing the system's field of view (FOV) can accommodate RCP components with $l = 2m(60)$. f, Intensity line profiles along the white dashed lines in d and e, quantifying undesired polarization leakage into the STED$_{xy}$ mode. The residual undesired polarization components for $m$ = 3, 5, and 30 are 36%, 18.8%, 1.3%, respectively. For $m$ = 1, polarization modes are not separable, resulting in a severely distorted depletion field. For $m$ = 2, STED$_{xy}$ and undesired components remain overlapped, making energy quantification infeasible.

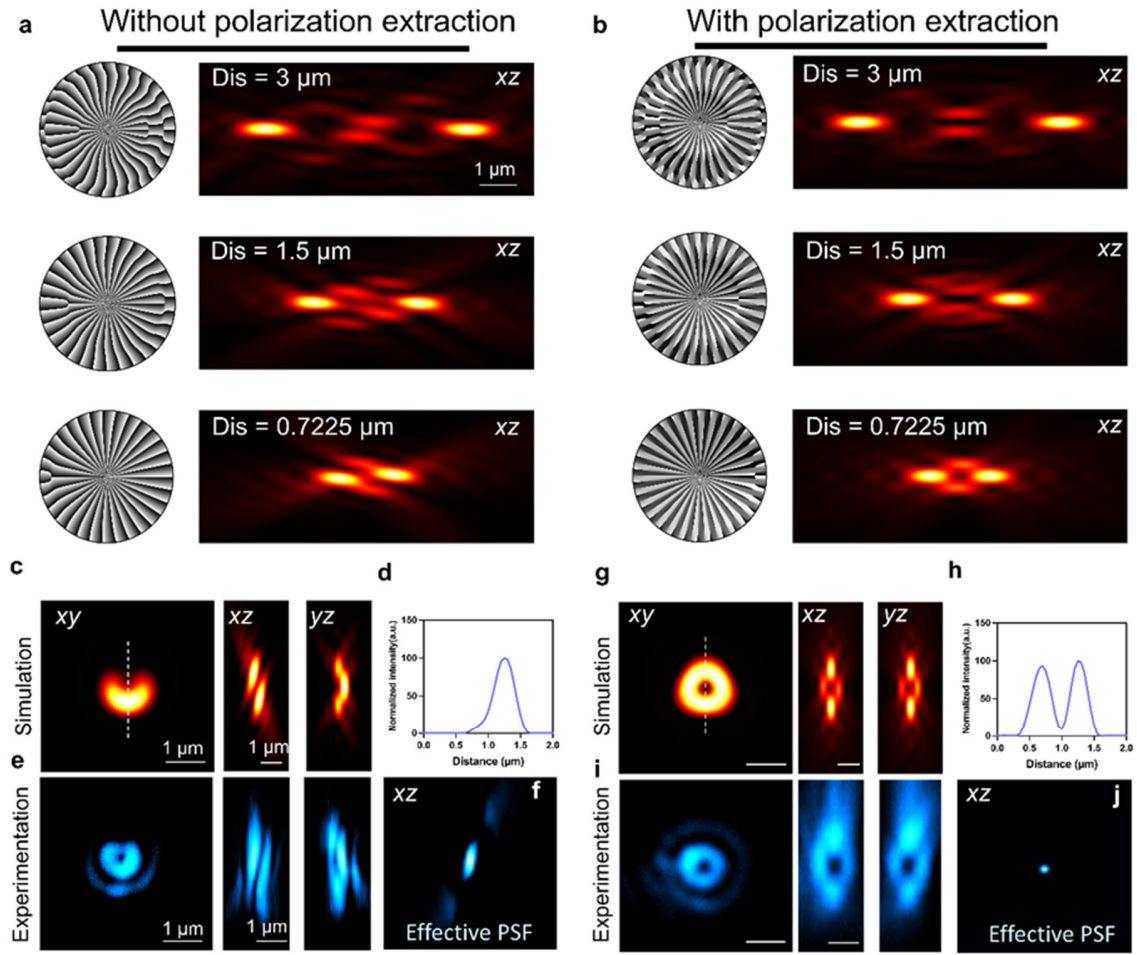

**Supplementary Figure 7** Impact of polarization extraction on depletion beam quality. a–b, Simulated $x$–$z$ cross-sections of the depletion light field without (a) and with (b) polarization separation between $STED_{xy}$ and $STED_z$. Three axial separations between foci ($\Delta z = 3\,\mu m$, $1.5\,\mu m$, and $0.7225\,\mu m$) are shown for each condition, All the phase parameters setting are shown in Supplementary Note. 3. Scale bar, $1\,\mu m$. c, g, Simulated depletion PSFs without (c) and with (g) polarization extraction, showing $x$–$y$, $x$–$z$, and $y$–$z$ views from left to right. Scale bar, $1\,\mu m$. d, h, Intensity line profiles along the dashed lines indicated in (c) and (g), respectively, demonstrating the suppression of interference artifacts with polarization separation. e, i, experimental depletion PSFs without (e) and with (i) polarization separation, visualized in $x$–$y$, $x$–$z$, and $y$–$z$ projections (left to right). Scale bar, $1\,\mu m$. f, j, simulated effective PSFs incorporating the STED beam patterns from (c) and (g), respectively, demonstrating isotropic compression only under orthogonal polarization conditions.

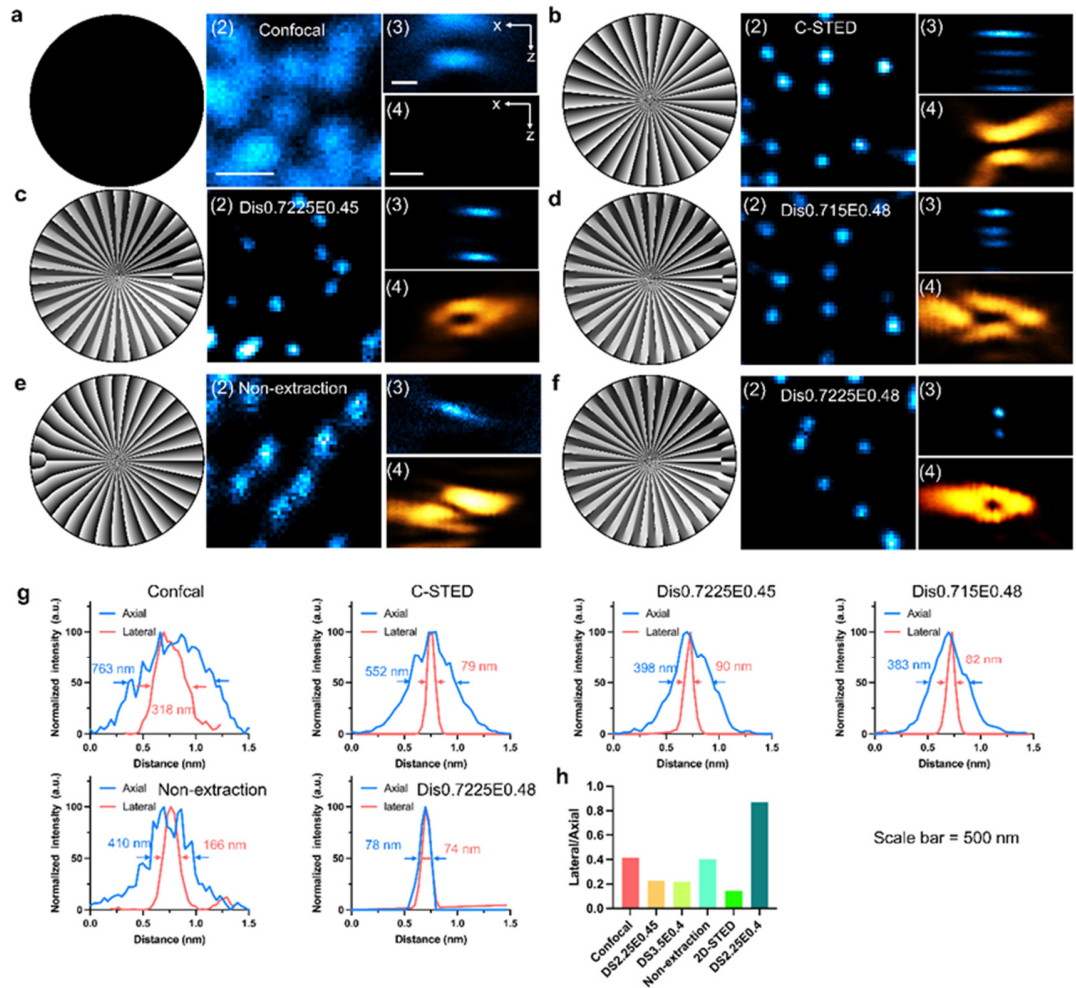

**Supplementary Figure 8** Modulation lateral and axial resolution of ISO-STED. a, Control condition without depletion beam. (1) No phase applied to the SLM. (2) Confocal $x$–$y$ image of 20 nm fluorescent beads. (3) Confocal $x$–$z$ cross-section of the same beads. (4) Depletion beam profile visualized by 80 nm gold nanoparticle scattering. b, C-STED configuration, with corresponding $x$–$y$, $x$–$z$, and depletion beam profiles analogous to a. c, ISO-STED configuration with STED$_{xy}$–STED$_z$ axial separation $\Delta z = 0.7225$ μm and energy ratio $s_{xy} = 0.45$, $s_z = 0.225$. d, ISO-STED with $\Delta z = 0.715$ μm and $s_{xy} = 0.48$, $s_z = 0.45$. e, ISO-STED without polarization separation. f, ISO-STED with $\Delta z = 0.7225$ μm and $s_{xy} = s_{xy} = 0.48$, $s_z = 0.45$. Scale bars, 500 nm, All the phase parameters setting are shown in Supplementary Note. 3. g, FWHM measurements of lateral and axial resolution under each condition: confocal, C-STED, Dis0.7E0.45 (c), Dis2E0.3 (d), non-extraction (e), and Dis0.7E0.3 (f). h, Corresponding ratio of axial to lateral FWHM for each condition; values approaching 1 indicate improved isotropy. All the phase parameters setting are shown in Supplementary Note. 3.

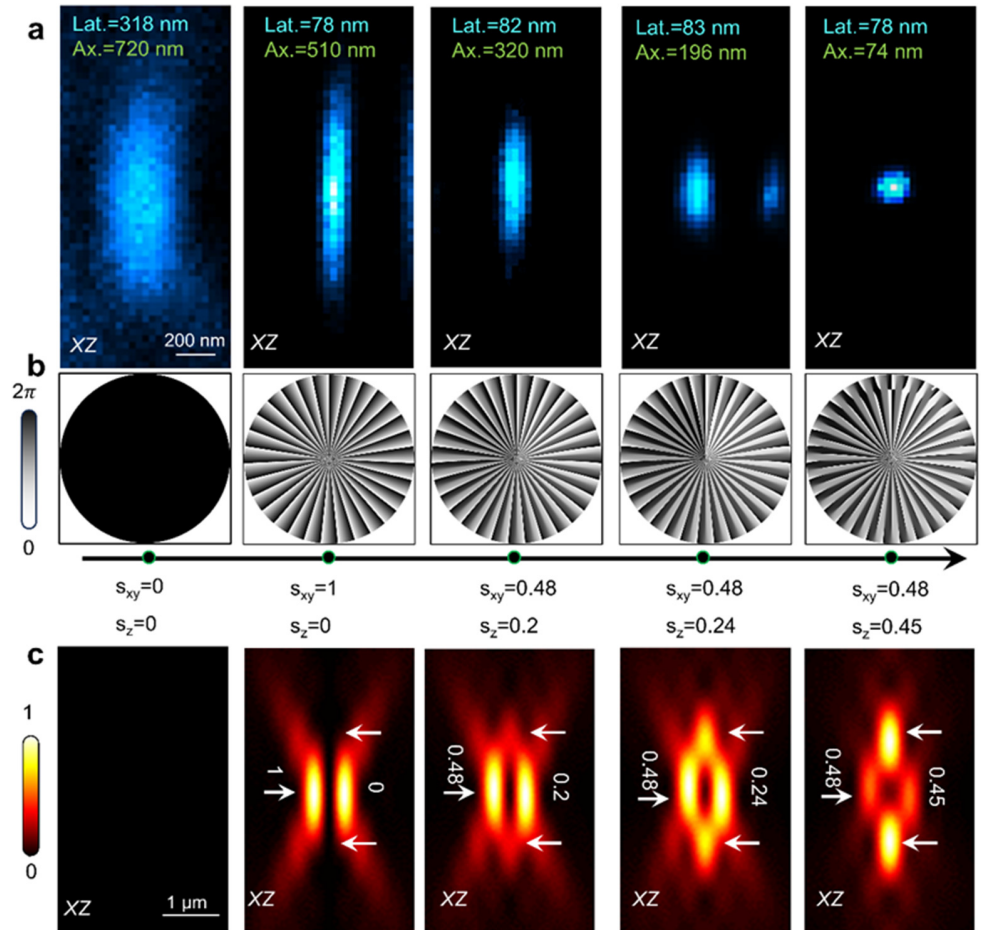

**Supplementary Figure 9** Optimization of 3D super-resolution in ISO-STED via STEDxy energy tuning. a, Experimental *x–z* images of 20 nm fluorescent beads acquired under ISO-STED configurations with varying STEDxy energy ratios ($s_z$= 0, 0.2, 0.24, 0.45). Scale bar, 200 nm. b, Corresponding SLM phase masks applied for each condition in a. c, Simulated depletion beam profiles (*x–z* view) for the same configurations shown in a, demonstrating the effect of STED$_{xy}$ energy modulation on axial confinement. Scale bar, 1 µm. All the phase parameters setting are shown in Supplementary Note. 3.

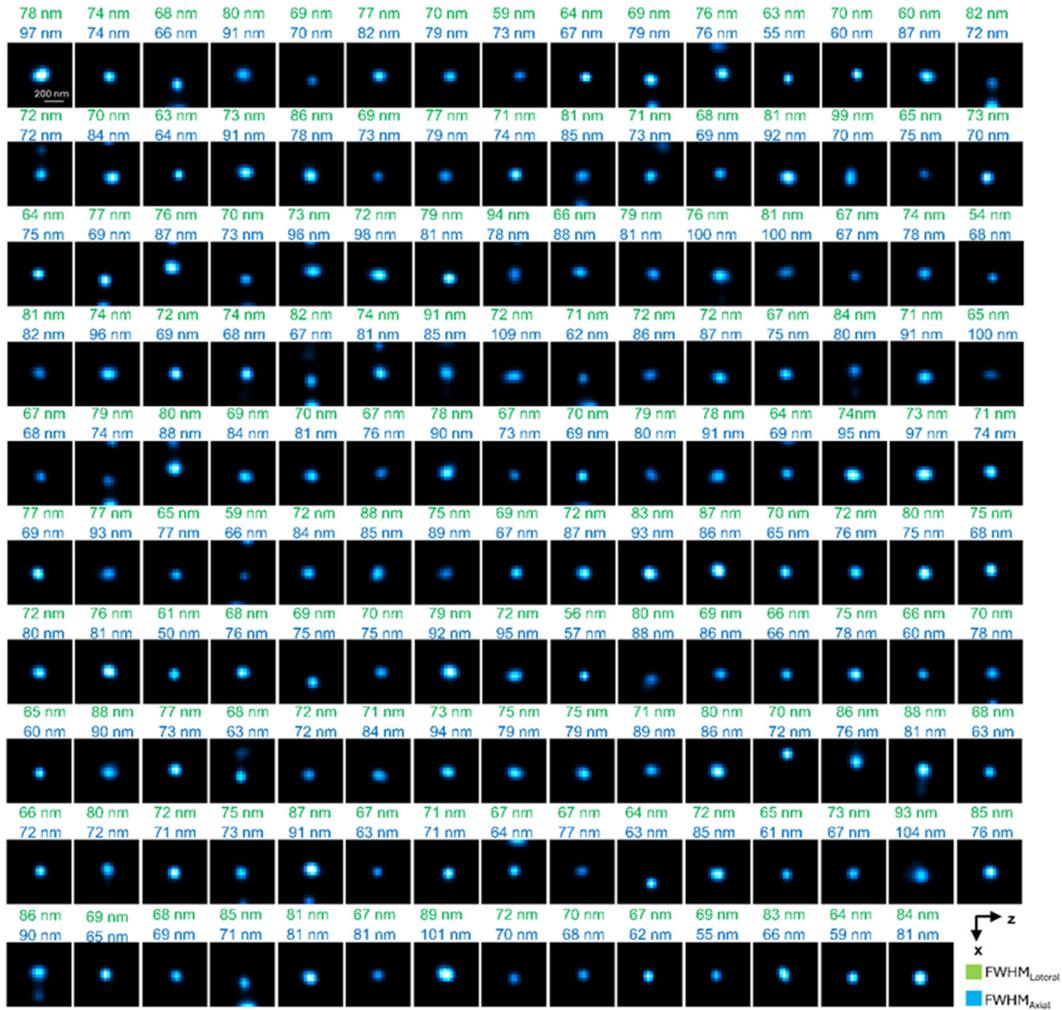

**Supplementary Figure 10** Quantification of 3D resolution. Crimson fluorescent beads (20 nm diameter) were imaged using ISO-STED nanoscopy. One-dimensional Lorentzian functions were fitted to intensity profiles of isolated beads along the x (lateral) and z (axial) axes. Full width at half maximum (FWHM) values extracted from the fits are labeled above each image in green (lateral) and blue (axial). All intensity values are normalized to their respective peaks.

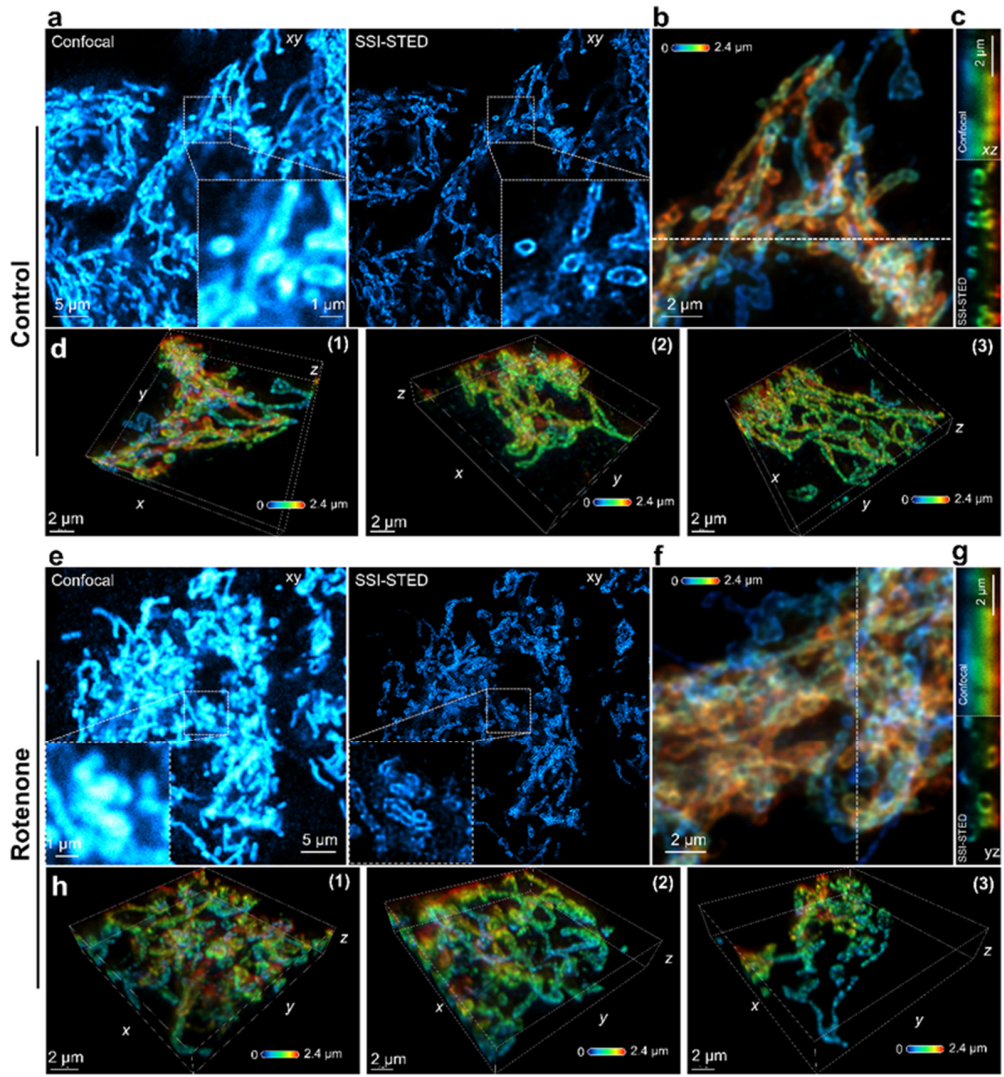

**Supplementary Figure 11** 3D reconstruction of mitochondrial morphology under oxidative stress using ISO-STED. a, e, Representative confocal and ISO-STED lateral-view images of mitochondria in control (a) and rotenone-treated (e) HeLa cells. White dashed boxes indicate magnified regions shown at bottom right. Scale bars: 5 μm (overview), 1 μm (zoom). b, f, 3D maximum intensity projections of ISO-STED volumes in control (b) and rotenone-treated (f) groups. Depth information is color-coded as indicated by the color bar. Scale bars: 2 μm. c, g, x–z cross-sections of confocal and ISO-STED data along the dashed lines in b and f. Scale bars: 2 μm. d, h, Representative 3D renderings of mitochondrial networks from three independent fields of view in control (d) and rotenone-treated (h) groups, highlighting altered morphology under oxidative stress. Scale bars: 2 μm.

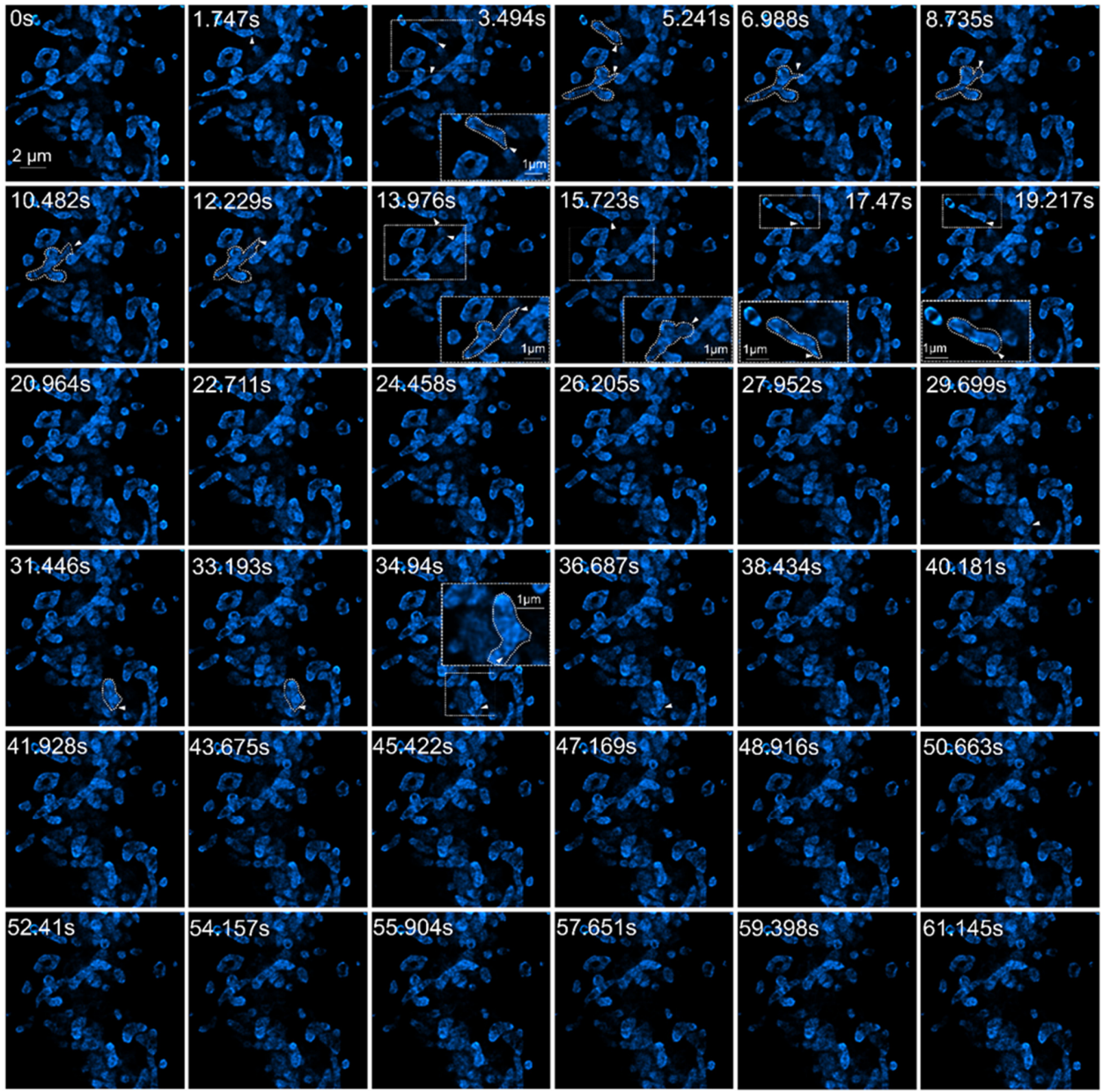

**Supplementary Figure 12** Time-lapse imaging of mitochondrial dynamics in live cells using ISO-STED. Inner mitochondrial membranes were imaged in live HeLa cells using ISO-STED at 1.747 s intervals over a total duration of 61.145 s. Regions of interest, highlighted by white arrows, indicate transient morphological events, such as sudden extension or retraction of mitochondrial tips. Insets show magnified views of these dynamic changes. These results demonstrate that the high axial resolution of ISO-STED, achieved via suppression of out-of-focus background, enables high-contrast monitoring of mitochondrial dynamics in live cells. scale bar, 2 μm.

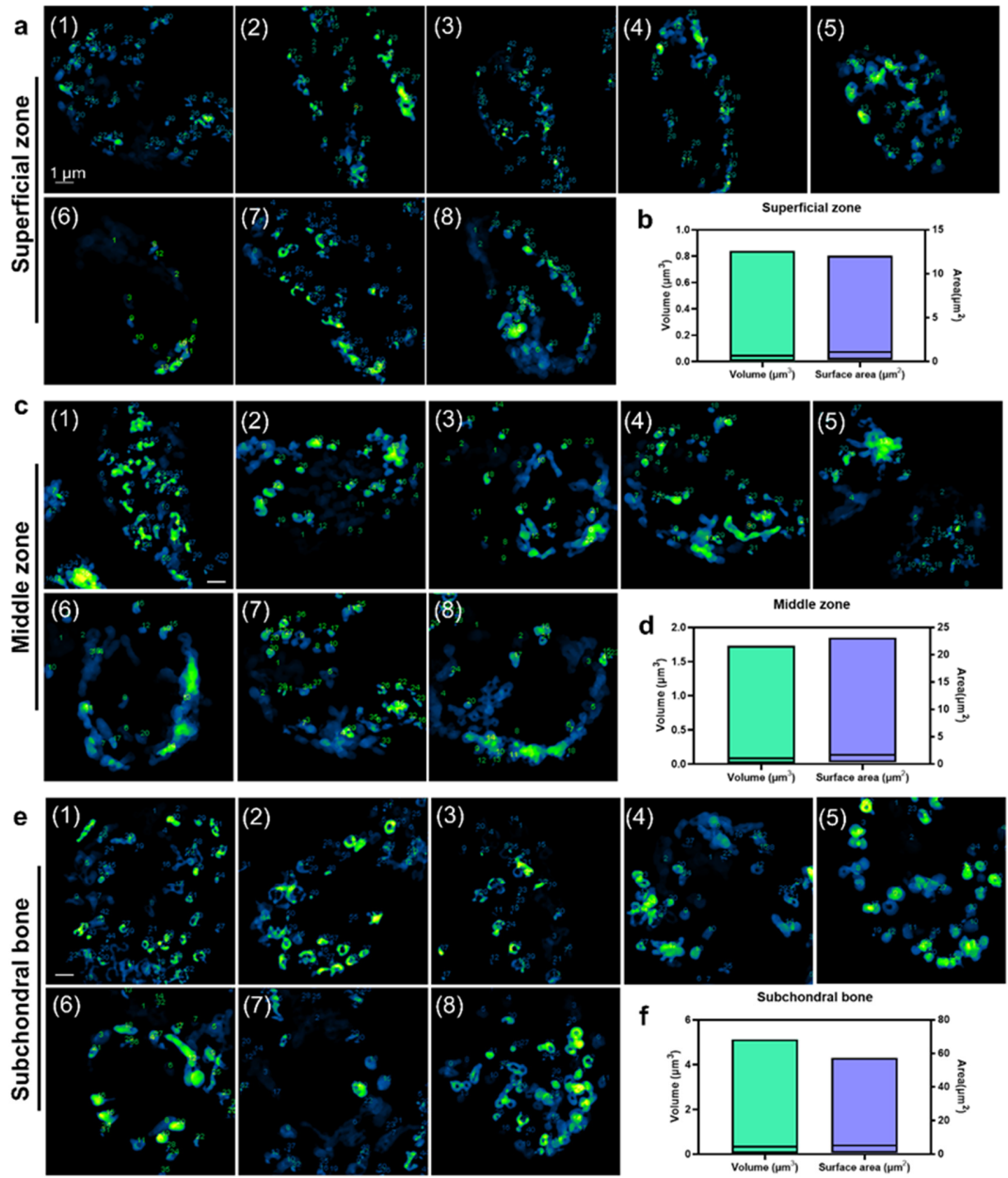

**Supplementary Figure 13** Quantitative analysis of mitochondrial volume across cartilage zones. a,c,e, Individual mitochondrial volumes from eight cells each in the superficial zone (a), middle zone (c), and subchondral bone zone (e), measured using the 3D Objects Counter plugin in FIJI. Numbers denote distinct mitochondrial objects within each field. b,d,f, Statistical summaries of mitochondrial size parameters in the respective zones. Left y-axis: mitochondrial volume (mean ± s.d.): superficial zone, 0.08447 μm³; middle zone, 0.1245 μm³; subchondral bone zone, 0.5576 μm³. Right y-axis: surface area (mean ± s.d.): superficial zone, 1.658 μm²; middle zone, 2.209 μm²; subchondral bone zone, 7.394 μm². These data reveal a marked increase in mitochondrial size in subchondral bone cells compared to more superficial layers.

| Figure | Imaging mode | Sample/structure | STED laser power (mW) | Volume size(W×H×D) (z step size or time interval) |
|---|---|---|---|---|
| **Supplementary Table 1:** Data acquisition parameters. | | | | |
| ×H×D×C represents voxel size in x, y, z dimensions | | | | |
| Figure 1d | Confocal | 20 nm deep-red beads (635nm) | 0 | 256×256×120 (25 nm) |
| | C-STED | 20 nm deep-red beads (635nm) | 50 | 256×256×120 (25 nm) |
| | ISO-STED | 20 nm deep-red beads (635nm) | 80 | 256×256×120 (25 nm) |
| Figure 2 a | Confocal | Fixed Hela cell β-tubulin | 0 | 1024×1024×1 |
| | C-STED | Fixed Hela cell β-tubulin | 50 | 1024×1024×1 |
| | ISO-STED | Fixed Hela cell β-tubulin | 80 | 1024×1024×1 |
| Figure 2 e | Confocal | Fixed Hela cell β-tubulin | 0 | 256×256×40 (50 nm) |
| | C-STED | Fixed Hela cell β-tubulin | 50 | 256×256×40 (50 nm) |
| | ISO-STED | Fixed Hela cell β-tubulin | 80 | 256×256×40 (50 nm) |
| Figure 2 g | Confocal | Fixed Hela cell nup153 | 0 | 512×512×60 (50 nm) |
| | C-STED | Fixed Hela cell nup153 | 50 | 512×512×60 (50 nm) |
| | ISO-STED | Fixed Hela cell nup153 | 80 | 512×512×60 (50 nm) |
| Figure 2 k | Confocal | Fixed Hela cell nup153 | 0 | 50×50×1 |
| | C-STED | Fixed Hela cell nup153 | 50 | 50×50×1 |
| | ISO-STED | Fixed Hela cell nup153 | 80 | 50×50×1 |
| Figure 3 a | Confocal | Live Hela cell IMM | 0 | 1024×1024×1 |
| | C-STED | Live Hela cell IMM | 50 | 1024×1024×1 |
| | ISO-STED | Live Hela cell IMM | 80 | 1024×1024×1 |
| Figure 3 d | Confocal | Live Hela cell IMM | 0 | 256×256×41 (50 nm) |
| | ISO-STED | Live Hela cell IMM | 80 | 256×256×41 (50 nm) |
| Figure 3 f, and Supplementary Figure 12 | ISO-STED | Live Hela cell IMM | 80 | 512×512×1 (1.57s time interval) |
| Figure 4 a | Confocal | Fixed mice cartilage tissue OMM | 0 | 512×512×100 (300 nm) |
| Figure 4 b-e | ISO-STED | Fixed mice cartilage tissue OMM | 80 | 256×256×60 (40 nm) |
| Figure 4 h | Confocal | Fixed mice brain tissue neuron | 0 | 512×512×75 (400 nm) |
| Figure 4 i-l | ISO-STED | Fixed mice brain tissue neuron | 85 | 256×256×60 (40 nm) |
| Figure 5 a | Bright field | H&E stained Fixed mice cartilage tissue | 0 | 1600×1200×1 |
| Figure 5 b(1),c(1),d(1) | Confocal | Fixed mice cartilage tissue mitochondria | 0 | 512×512×100 (300 nm) |
| Figure 5 b(2),c(3),d(4) | ISO-STED | Fixed mice cartilage tissue mitochondria | 80 | 256×256×60 (40 nm) |
| Supplementary Figure 10 | All modality | 20 nm deep-red beads (635nm) | 80 (Confocal:0) | 256×256×120 (25 nm) |
| Supplementary Figure 11 (a),(e) left panel | Confocal | Fixed Hela cell OMM | 0 | 1024×1024×1 |
| Supplementary Figure 11 (a),(e) right panel | ISO-STED | Fixed Hela cell OMM | 85 | 1024×1024×1 |
| Supplementary Figure 11 b-d, f-h | ISO-STED | Fixed Hela cell OMM | 85 | 256×256×60 (40 nm) |
| Supplementary Figure 13 | ISO-STED | Fixed mice cartilage tissue mitochondria | 80 | 256×256×60 (40 nm) |

# Supplementary Note 1: Principle of optical pen and polarization mode extraction

Achieving a truly isotropic three-dimensional (3D) depletion focus in STED nanoscopy requires the generation of multiple foci along the axial (z) direction, with precise control over their spatial positions, relative intensities, orbital angular momentum (OAM), and polarization states. The ability to assign orthogonal polarization states to each focus is particularly critical, as it suppresses mutual interference that would otherwise distort the depletion field. To this end, we combine two key strategies in our ISO-STED implementation: an optical pen technique for multiplexed focal generation and polarization mode extraction via a vortex phase retarder[1,2]. This note details the theoretical foundation and implementation of both approaches.

## Polarization Mode Extraction

Linearly polarized light can be represented as a coherent superposition of left- and right-circularly polarized components. These components are mutually orthogonal and typically maintain a stable, intertwined polarization state during propagation. Mathematically, the electric field of a linearly polarized beam carrying OAM can be expressed as:

$$E = \exp\left(imf\left(\varphi,\theta\right)\right)\left|\mathbf{R}\right\rangle + \exp\left(-imf\left(\varphi,\theta\right)\right)\left|\mathbf{L}\right\rangle \tag{1}$$

, where $\left|\mathbf{R}\right\rangle = \left[1,-i\right]$, $\left|\mathbf{L}\right\rangle = \left[1,i\right]$ denote right- and left-circularly polarized states, respectively; and $f\left(\varphi,\theta\right) = \varphi$ represents a spatially varying phase function that depends on the azimuthal angle $\varphi$ and the convergence angle $\theta$.

To break the inherent polarization symmetry and isolate the circular components, we introduce a high-order vortex phase retarder that imparts additional OAM-dependent phase shifts. Upon passing through the vortex retarder $P_{vortex}\left(\varphi\right) = \exp\left(\pm in\varphi\right)$, the polarization state is transformed as:

$$E_{mcL} = \exp\left(i\left(m+n\right)\varphi\right)\left|\mathbf{R}\right\rangle + \exp\left(-i\left(m-n\right)\varphi\right)\left|\mathbf{L}\right\rangle, \tag{2}$$

$$E_{mcR} = \exp\left(-i\left(m+n\right)\varphi\right)\left|\mathbf{L}\right\rangle + \exp\left(i\left(m-n\right)\varphi\right)\left|\mathbf{R}\right\rangle \tag{3}$$

This result shows that the $\left|\mathbf{R}\right\rangle$ now carries a topological charge of $m+n$, while the $\left|\mathbf{L}\right\rangle$ component carries $m-n$. When $n=m$, Eq.(2) and (3) turn to $E_{mcL} = \exp\left(i2m\varphi\right)\left|\mathbf{R}\right\rangle + \left|\mathbf{L}\right\rangle$, and $E_{mcL} = \exp\left(-i2m\varphi\right)\left|\mathbf{L}\right\rangle + \left|\mathbf{R}\right\rangle$. The large difference in OAM enables spatial separation of the two

polarization modes, effectively decoupling their originally entangled states. Since $f(\varphi, \theta) = \varphi$ is spatially dependent, the output beam becomes non-uniformly polarized, breaking the symmetry of the input beam. This spatial separation forms the foundation of our polarization mode extraction strategy, allowing dynamic and selective control over individual foci based on their polarization and OAM.

## Optical Pen for Multiplexed Foci Generation

To generate multiple tightly focused beams in 3D with tunable spatial positions, energy, and OAM, we extend the optical pen framework based on the light field distribution near the focal region. After focusing through a high-NA objective, the optical path difference $\Delta s$ between rays forming off-axis foci ($AP - BP$) can be expressed as[1]:

$$\Delta s = \rho \sin\theta \cos(\varphi - \varphi_p) + z(1 - \cos\theta) \tag{4}$$

Here, $\rho$ is the radial distance in spherical coordinates, $\theta$ and $\varphi$ represent the convergence angle and azimuthal angle, and $z$ denotes the axial displacement. To generate a focus at this off-axis position, the incoming beam must carry a compensating phase term:

$$\psi = k\Delta s \tag{5}$$

, where $k = 2\pi n / \lambda$ is the wave number，$n$ is the refractive index and $\lambda$ is the wavelength. The corresponding pupil function for a single focus becomes $T = \exp(i\psi)$, and the OAM modulation component can be integrated with this function: $T_{OAM} = \exp(im\varphi)$, $m$ is the topology number. When $N$ foci are desired, the pupil function becomes a superposition:

$$T = \sum_{j=1}^{N} \left( s_j \exp(i\psi_j) \exp(im_j\varphi) \right) \tag{6}$$

, where $s_j$ modulates the intensity of each focus and $\psi_j$ encodes its spatial position and phase. In practical implementations, amplitude modulation suffers from low efficiency; thus, we extract only the phase term for SLM encoding:

$$T = \exp\left\{ i\left[ Phase\left( \sum_{j=1}^{N} s_j \exp(i\psi_j + \sigma_j) \exp(im_j\varphi) \right) \right] \right\} \tag{7}$$

Here, we use the above equation as the pupil filtering of the optical pen. For more convenient representation, this filtering function can be expressed as:

$$T = \exp\left\{ i \left[ Phase\left( \sum_{j=1}^{N} PF(s_j, x_j, y_j, z_j, \sigma_j, m_j) \right) \right] \right\} \tag{8}$$

This formulation enables full control over the number, energy ($s_j$), 3D positions ($x_j, y_j, z_j$), and OAM ($m_j$) of the generated foci (**Supplementary Fig.4**). In our configuration, we generate two axially offset foci at $z = \pm\Delta z$, each with topological charge $m$, and a third focus at $z = 0$ with topological charge –(m+1). After polarization mode extraction via the vortex retarder (Eq. 1–3), the two axially displaced foci (STED$_z$) carry right-circular polarization, while the central transverse focus (STED$_{xy}$) is left-circularly polarized, which avoid the interference among them. By optimizing the axial separation ($\Delta z$) and intensity coefficients ($s_j$), we achieve a uniformly distributed, isotropic 3D depletion light field.

When this depletion PSF is temporally and spatially overlapped with the excitation beam, it induces stimulated emission depletion across all directions, yielding a highly isotropic point spread function suitable for 3D super-resolution imaging.

## Supplementary Note 2: Simulation of ISO-STED depletion beam

This note presents the theoretical foundation for simulating the ISO-STED depletion beam, including the complete derivation of the electric field distribution near the focal region. As illustrated in Supplementary Figure 3, the depletion beam in the ISO-STED system is generated by focusing the SLM-modulated light field through a high-NA objective lens. The focal field distribution is computed based on the vectorial Debye diffraction integral.

According to Debye theory, the electric field $E(P)$ at an arbitrary point $P$ near the focus is given by:

$$E(r,\varphi,z) = -\frac{ik}{2\pi} \iint_{\Omega} a_1(\theta,\varphi) \exp(iks \bullet \rho) d\Omega \tag{9}$$

Here, $a_1(\theta,\varphi)$ is the vectorial amplitude on the spherical wavefront, $s$ is the unit vector in the propagation direction after refraction, and $\rho$ is the displacement vector from the geometric focus to the point $P$. When the incident light passes through the entrance pupil of the objective, the field at a radial coordinate $\rho$ on the pupil plane is described by:：

$$r_p = f \bullet g(\theta) \tag{10}$$

, where $g(\theta)$ is the incident function, it describes how light enters the lens plane and then the distribution of the wavefront after it is converged by the lens. For an aplanatic objective satisfying the sine condition：$g(\theta) = \sin\theta$，By substituting it into Eq.(10), we can get $r_p = f\sin\theta$，and $l(\theta) = J_1\left(2\beta_0(\sin\theta/\sin\alpha)\right)\exp\left(-\left(\beta_0(\sin\theta/\sin\alpha)\right)^2\right)$ [3]，($J_1(\bullet)$ is the Bessel function of first kind with order 1.) present the Gaussian function amplitude distribution of incident beam，In cylindrical coordinates, $s \bullet \rho$ can be calculated as

$$s \bullet \rho = z\cos\theta + r\sin\theta\cos(\varphi - \varphi_p) \tag{11}$$

The unit surface for the spherical area integral of the wavefront can be expressed as: $d\Omega = \sin\theta d\theta d\varphi$，Moreover, taking into account the cutting-edge effect of the lens, the expression of can be:

$$a_1(\theta,\varphi) = f\sqrt{\cos\theta} \bullet l(\theta) \bullet \mathbf{V} \tag{12}$$

Here, $\mathbf{V}$ represents the distribution of polarization.。By substituting $a_1(\theta,\varphi)$，$s \bullet \rho$, and into

Debye equation (Eq.8), we can obtain:

$$E(r,\varphi,z) = -\frac{ikf}{2\pi} \int_0^\alpha \int_0^{2\pi} \sin\theta \sqrt{\cos\theta} l(\theta) T\mathbf{V} \exp\left(ik\left(z\cos\theta + r\sin\theta\cos\left(\varphi-\varphi_p\right)\right)\right) d\varphi d\theta \quad (13)$$

Here, $\alpha = \arcsin(NA/n)$ is the maximum convergence angle defined by the numerical aperture ($NA$) and refractive index $n$ of the immersion medium. For $\mathbf{V}$, Based on Eqs. (2) and (3) of Supplementary Note 1, an input vortex beam (VB) with topological charge $m$ can be decomposed into two orthogonal circular polarization components: $E_{VB} = \exp(im\varphi)|\mathbf{R}\rangle + \exp(-im\varphi)|\mathbf{L}\rangle$, These components can be further projected onto linearly polarized x- and y-axes as:

$$E_{VB} = \exp(im\varphi) \bullet \left(|x\rangle + i|y\rangle\right) + \exp(-im\varphi) \bullet \left(|x\rangle - i|y\rangle\right) \quad (14)$$

, where $|x\rangle$ and $|y\rangle$ are the x and y linear polarization.

Upon passing through a vortex retarder, the polarization states acquire distinct topological charges (e.g., $2m$ for the right-circularly polarized component and 0 for the left-circularly polarized one), enabling selective polarization mode extraction. When $m = 30$, the power density of the unwanted polarization mode becomes negligible and can be ignored in practice. Thus, the electric field vector of the resulting beam can be described by either:

$$\mathbf{V} = \mathbf{V}_x + i\mathbf{V}_y \text{ or } \mathbf{V} = \mathbf{V}_x - i\mathbf{V}_y \quad (15)$$

, where $\mathbf{V}_x$ 和 $\mathbf{V}_y$ are the electric vectors of $|x\rangle$ and $|y\rangle$, respectively, and can be presented as[4]:

$$\mathbf{V}_x = \begin{bmatrix} \cos\theta\cos^2\varphi + \sin^2\varphi \\ -(1-\cos\theta)\sin\varphi\cos\varphi \\ -\sin\theta\cos\varphi \end{bmatrix}; \mathbf{V}_y = \begin{bmatrix} -(1-\cos\theta)\sin\varphi\cos\varphi \\ (\cos\theta)\sin^2\varphi + \cos^2\varphi \\ -\sin\theta\cos\varphi \end{bmatrix} \quad (16)$$

Substituting Eq. (15) and the spatial phase modulation from Eq. (7) of Supplementary Note 1 into the Debye function (Eq. 13), we can simulate the full 3D intensity distribution of the depletion beam, and the intensity distribution $I = |E|^2$. By tuning the spatial parameters, polarization states, and OAM for each focus, this framework enables the generation of a depletion PSF with isotropic suppression capability in 3D. When temporally and spatially overlapped with the excitation focus, this depletion field induces uniform stimulated emission depletion in all directions, yielding an isotropic PSF suitable for 3D super-resolution imaging.

# Supplementary Note 3: Parameters of all phases

## Figure 1 and supplementary Figure 1

Phase of the Figure 1 b:

From left to right: $T_{STEDz1} = PF(0.45, 0, 0, -0.7225, 0, 30)$ ; $T_{STEDz2} = PF(0.45, 0, 0, 0.7225, 0, 30)$ ;

$T_{STEDxy} = PF(0.48, 0, 0, 0, 0, -31)$

Phase of supplementary Figure 1(b-I):

$T_{b-I} = PF(1, 0, 0, 0, 0, 0)$ ;

Phase of supplementary Figure 1(b-II):

$T_{b-II} = \sum_{j=1}^{N} PF(s_j, x_j, y_j, z_j, \sigma_j, m_j)$ , where $x_{1-3} = 0; y_{1-3} = 0;$ $z_1 = -0.7225; z_2 = 0; z_3 = 0.7225;$

$s_{1,3} = 0.45; s_2 = 0.48; m_{1,3} = 30, m_2 = 30$ .

Phase of supplementary Figure 4:

$T_a = PF(1, 0, 0, 0, 0, 0)$ ;

$T_b = PF(s_i, 0, 0, z_i, 0, m_i)$ , where $s_{1-2} = 0.5$ ; $z_1 = -2; z_2 = 2$ ; $m_{1-2} = 30$ .

$T_c = PF(s_i, 0, 0, z_i, 0, m_i)$ , where $s_{1-3} = 0.33$ ; $z_1 = -4; z_2 = 0; z_3 = 4$ ; $m_{1,3} = 30; m_2 = -31$ .

$T_d = PF(s_i, 0, 0, z_i, 0, m_i)$ , where $s_{1-3} = 0.33$ ; $z_1 = -6; z_2 = 0; z_3 = 6$ ; $m_{1,3} = 30; m_2 = -31$ .

$T_e = PF(s_i, 0, 0, z_i, 0, m_i)$ , where $s_1 = 0.4; s_2 = 0.4; s_3 = 0.2$ ; $z_1 = -4; z_2 = 0; z_3 = 4$ ; $m_{1,3} = 30; m_2 = -31$ .

$T_e = PF(s_i, 0, 0, z_i, 0, m_i)$ , where $s_1 = 0.45; s_2 = 0.48; s_3 = 0.45$ ; $z_1 = -0.7225; z_2 = 0; z_3 = 0.7225$ ;

$m_{1,3} = 30; m_2 = -31$ .

Phase of supplementary Figure 5:

$T = PF(s_i, 0, 0, z_i, 0, m_i)$ , where $s_{1-3} = 0.33$ ; $z_1 = -6; z_2 = 0; z_3 = 6$ ; $m_{1,3} = 30; m_2 = -31$ .

Phase of supplementary Figure 6(a):

From left to right: $T_1 = \exp(i\varphi)$ ; $T_2 = \exp(i2\varphi)$ ; $T_3 = \exp(i3\varphi)$ ; $T_4 = \exp(i5\varphi)$ ; $T_5 = \exp(i30\varphi)$ .

Phase of supplementary Figure 6(d):

From left to right:

$T_1 = PF(s_i, 0, 0, z_i, 0, m_i)$ , where $s_1 = s_3 = 0.45$ ; $s_1 = 0.48$ ; $z_1 = -0.7225; z_2 = 0; z_3 = 0.7225$ ;

$m_{1,3} = 1; m_2 = 0$ .

$T_2 = PF(s_i, 0, 0, z_i, 0, m_i)$ , where $s_1 = s_3 = 0.45$ ; $s_1 = 0.48$ ; $z_1 = -0.7225; z_2 = 0; z_3 = 0.7225$ ; $m_{1,3} = 2; m_2 = -1$ .

$T_3 = PF(s_i, 0, 0, z_i, 0, m_i)$ , where $s_1 = s_3 = 0.45$ ; $s_1 = 0.48$ ; $z_1 = -0.7225; z_2 = 0; z_3 = 0.7225$ ; $m_{1,3} = 3; m_2 = -4$ .

$T_5 = PF(s_i, 0, 0, z_i, 0, m_i)$ , where $s_1 = s_3 = 0.45$ ; $s_1 = 0.48$ ; $z_1 = -0.7225; z_2 = 0; z_3 = 0.7225$ ; $m_{1,3} = 5; m_2 = -6$ .

$T_6 = PF(s_i, 0, 0, z_i, 0, m_i)$ , where $s_1 = s_3 = 0.45$ ; $s_1 = 0.48$ ; $z_1 = -0.7225; z_2 = 0; z_3 = 0.7225$ ; $m_{1,3} = 30; m_2 = -31$ .

Phase of supplementary Figure 7:

Without polarization extraction (from top to bottom):

$T_1 = PF(s_i, 0, 0, z_i, 0, m_i)$, where $s_{1,3} = 0.45; s_2 = 0.48$ ; $z_1 = -3; z_2 = 0; z_3 = 3$ ; $m_{1,3} = 30; m_2 = 31$ .

$T_2 = PF(s_i, 0, 0, z_i, 0, m_i)$, where $s_{1,3} = 0.45; s_2 = 0.48$ ; $z_1 = -1.5; z_2 = 0; z_3 = 1.5$ ; $m_{1,3} = 30; m_2 = 31$ .

$T_3 = PF(s_i, 0, 0, z_i, 0, m_i)$ , where $s_{1,3} = 0.45; s_2 = 0.48$ ; $z_1 = -0.7225; z_2 = 0; z_3 = 0.7225$ ; $m_{1,3} = 30; m_2 = 31$ .

With polarization extraction (from top to bottom):

$T_1 = PF(s_i, 0, 0, z_i, 0, m_i)$, where $s_{1,3} = 0.45; s_2 = 0.48$ ; $z_1 = -3; z_2 = 0; z_3 = 3$ ; $m_{1,3} = 30; m_2 = -31$ .

$T_2 = PF(s_i, 0, 0, z_i, 0, m_i)$, where $s_{1,3} = 0.45; s_2 = 0.48$ ; $z_1 = -1.5; z_2 = 0; z_3 = 1.5$ ; $m_{1,3} = 30; m_2 = -31$ .

$T_3 = PF(s_i, 0, 0, z_i, 0, m_i)$ , where $s_{1,3} = 0.45; s_2 = 0.48$ ; $z_1 = -0.7225; z_2 = 0; z_3 = 0.7225$ ; $m_{1,3} = 30; m_2 = -31$ .

Phase of supplementary Figure 8:

$T_a = PF(1, 0, 0, 0, 0, 0)$ ;

$T_b = PF(1, 0, 0, 0, 0, -31)$ ;

$T_c = PF(s_i, 0, 0, z_i, 0, m_i)$ , where $s_{1,3} = 0.275; s_2 = 0.45$ ; $z_1 = -0.7225; z_2 = 0; z_3 = 0.7225$ ; $m_{1,3} = 30; m_2 = -31$ .

$T_d = PF(s_i, 0, 0, z_i, 0, m_i)$ , where $s_{1,3} = 0.45; s_2 = 0.48,$ ; $z_1 = -0.715; z_2 = 0; z_3 = 0.715$ ; $m_{1,3} = 30; m_2 = -31$ .

$T_e = PF(s_i, 0, 0, z_i, 0, m_i)$ , where $s_{1,3} = 0.45; s_2 = 0.48,$ ; $z_1 = -0.7225; z_2 = 0; z_3 = 0.7225$ ; $m_{1,3} = 30; m_2 = 31$ .

$T_f = PF(s_i, 0, 0, z_i, 0, m_i)$ , where $s_{1,3} = 0.45; s_2 = 0.48,$ ; $z_1 = -0.7225; z_2 = 0; z_3 = 0.7225$ ; $m_{1,3} = 30; m_2 = -31$ .

Phase of supplementary Figure 9:

From left to right in supplementary Figure 8(b):

$T_1 = PF(1, 0, 0, 0, 0, 0)$ ;

$T_2 = PF(1, 0, 0, 0, 0, -31)$ ;

$T_3 = PF(s_i, 0, 0, z_i, 0, m_i)$ , where $s_{1,3} = 0.2; s_2 = 0.48$ ; $z_1 = -0.7225; z_2 = 0; z_3 = 0.7225$ ; $m_{1,3} = 30; m_2 = -31$ .

$T_4 = PF(s_i, 0, 0, z_i, 0, m_i)$ , where $s_{1,3} = 0.24; s_2 = 0.48$ ; $z_1 = -0.7225; z_2 = 0; z_3 = 0.7225$ ; $m_{1,3} = 30; m_2 = -31$ .

$T_5 = PF(s_i, 0, 0, z_i, 0, m_i)$ , where $s_{1,3} = 0.45; s_2 = 0.48$ ; $z_1 = -0.7225; z_2 = 0; z_3 = 0.7225$ ; $m_{1,3} = 30; m_2 = -31$ .

**Supplementary References**

1.  Weng X., Song Q., Li X., et al. Free-space creation of ultralong anti-diffracting beam with multiple energy oscillations adjusted using optical pen. *Nat. Commun.* **9**, 5035 (2018).

2.  Weng X., Miao Y., Zhang Q., et al. Extraction of Inherent Polarization Modes from an m‑Order Vector Vortex Beam. *Adv. Photon. Res.* **3**, 10 (2022).

3.  Fuscaldo, W. et al. Bessel-Gauss Beams Through Leaky Waves: Focusing and Diffractive Properties. *Phys. Rev. Applied* **13**, 064040 (2020).

4.  Richards, B., Wolf, E. Electromagnetic Diffraction in Optical Systems. II. Structure of the Image Field in an Aplanatic System. *Proc. Royal Soc. A* **253**, 358-379 (1959).